\documentclass[12pt]{article}
\usepackage{amssymb}
\usepackage{graphics}

\DeclareFontFamily{U}{rsf}{}
\DeclareFontShape{U}{rsf}{m}{n}{
  <5> <6> rsfs5 <7> <8> <9> rsfs7 <10-> rsfs10}{}
\DeclareMathAlphabet\Scr{U}{rsf}{m}{n}

\catcode`\@=11
\def\@citex[#1]#2{%
\if@filesw \immediate \write \@auxout {\string \citation {#2}}\fi
\@tempcntb\m@ne \let\@h@ld\relax \def\@citea{}%
\@cite{%
  \@for \@citeb:=#2\do {%
    \@ifundefined {b@\@citeb}%
      {\@h@ld\@citea\@tempcntb\m@ne{\bf ?}%
      \@warning {Citation `\@citeb ' on page \thepage \space undefined}}%
      {\@tempcnta\@tempcntb \advance\@tempcnta\@ne%
      \@tempcntb\number\csname b@\@citeb \endcsname \relax%
      \ifnum\@tempcnta=\@tempcntb 
        \ifx\@h@ld\relax%
          \edef \@h@ld{\@citea\csname b@\@citeb\endcsname}%
        \else%
          \edef\@h@ld{\ifmmode{-}\else--\fi\csname b@\@citeb\endcsname}%
        \fi%
      \else
        \@h@ld\@citea\csname b@\@citeb \endcsname%
        \let\@h@ld\relax%
      \fi}%
    \def\@citea{,\penalty\@highpenalty\,}%
  }\@h@ld
}{#1}}

\def\@citeb#1#2{{[#1]\if@tempswa , #2\fi}}
\def\@citeu#1#2{{$^{#1}$\if@tempswa , #2\fi }}
\def\@citep#1#2{{#1\if@tempswa , #2\fi}}

\def\bcites{         
        \catcode`\@=11
        \let\@cite=\@citeb
        \catcode`\@=12
}

\def\upcites{         
        \catcode`\@=11
        \let\@cite=\@citeu
        \catcode`\@=12
}

\def\plaincites{      
        \catcode`\@=11
        \let\@cite=\@citep
        \catcode`\@=12
}

\newcount\hour
\newcount\minute
\newtoks\amorpm
\hour=\time\divide\hour by 60
\minute=\time{\multiply\hour by 60 \global\advance\minute by-\hour}
\edef\standardtime{{\ifnum\hour<12 \global\amorpm={am}%
        \else\global\amorpm={pm}\advance\hour by-12 \fi
        \ifnum\hour=0 \hour=12 \fi
        \number\hour:\ifnum\minute<10 0\fi\number\minute\the\amorpm}}
\edef\militarytime{\number\hour:\ifnum\minute<10 0\fi\number\minute}

\def\draftlabel#1{{\@bsphack\if@filesw {\let\thepage\relax
   \xdef\@gtempa{\write\@auxout{\string
      \newlabel{#1}{{\@currentlabel}{\thepage}}}}}\@gtempa
   \if@nobreak \ifvmode\nobreak\fi\fi\fi\@esphack}
        \gdef\@eqnlabel{#1}}
\def\@eqnlabel{}
\def\@vacuum{}
\def\marginnote#1{}
\def\draftmarginnote#1{\marginpar{\raggedright\scriptsize\tt#1}}
\def\draft{
        \pagestyle{plain}
        \overfullrule=2pt
        \oddsidemargin -.5truein
        \def\@oddhead{\sl \phantom{\today\quad\militarytime} \hfil
        \smash{\Large\sl DRAFT} \hfil \today\quad\militarytime}
        \let\@evenhead\@oddhead
        \let\label=\draftlabel
        \let\marginnote=\draftmarginnote
        \def\ps@empty{\let\@mkboth\@gobbletwo
        \def\@oddfoot{\hfil \smash{\Large\sl DRAFT} \hfil}
        \let\@evenfoot\@oddhead}
        \def\@eqnnum{(\theequation)\rlap{\kern\marginparsep\tt\@eqnlabel}%
        \global\let\@eqnlabel\@vacuum}  }

\def\section{\@startsection {section}{1}{\z@}{3.ex plus 1ex minus
 .2ex}{2.ex plus .2ex}{\large\bf}}
\def\subsection{\@startsection{subsection}{2}{\z@}{2.75ex plus 1ex minus
 .2ex}{1.5ex plus .2ex}{\bf}}        

\def\appendix{{\newpage\section*{Appendix}}\let\appendix\section%
        {\setcounter{section}{0}
        \gdef\thesection{\Alph{section}}}\section}

\def\abstract{\if@twocolumn
\section*{Abstract}
\else 
\begin{center}
{\bf Abstract\vspace{-.5em}\vspace{0pt}}
\end{center}
\quotation
\fi}

\catcode`\@=12

\newcommand{\beq}{\begin{equation}}
\newcommand{\eeq}{\end{equation}}
\newcommand{\beqa}{\begin{eqnarray}}
\newcommand{\eeqa}{\end{eqnarray}}
\newcommand{\dd}{{\rm d}}

\newcommand{\Z}{{\mathbb Z}}

\newcommand{\R}{{\mathbb R}}
\newcommand{\C}{{\mathbb C}}

\newcommand{\e}{\,{\rm e}}

\newcommand{\be}{\begin{equation}}
\newcommand{\ee}{\end{equation}}
\newcommand{\bea}{\begin{eqnarray}}
\newcommand{\eea}{\end{eqnarray}}

\def\to{\rightarrow}

\def\lae{\mathrel{\mathop{\smash{\lower .5 ex \hbox{$\stackrel<\sim$}}}}}
\def\lae{\mathrel{\mathop{\smash{\lower .5 ex \hbox{$\stackrel>\sim$}}}}}

\def\Tr{{\rm Tr}}
\def\l:{\mathopen{:}\,}
\def\r:{\,\mathclose{:}}

\catcode`\@=11
\def\theequation{\arabic{equation}}
\catcode`\@=12

\bcites

\catcode`\@=11
\def\theequation{\thesection.\arabic{equation}}
\@addtoreset{equation}{section}
\@addtoreset{footnote}{section}
\@addtoreset{footnote}{subsection}
\catcode`\@=12

\newcommand{\opsi}{\overline{\psi}}
\newcommand{\oQ}{\overline{Q}}
\newcommand{\btheta}{\overline{\theta}}

\newcommand{\bepsilon}{\overline{\epsilon}}

\newcommand{\bi}{\overline{\imath}}

\newcommand{\bareta}{\overline{\eta}}

\newcommand{\s}{\sigma}
\newcommand{\nn}{\nonumber}
\newcommand{\Hom}{{\rm Hom}}
\newcommand{\bartial}{\overline{\partial}}
\newcommand{\RR}{{}_{\rm RR}}
\newcommand{\NSNS}{{}_{\rm NSNS}}
\newcommand{\g}{\gamma}
\newcommand{\kket}[1]{\vert  #1\rangle\!\rangle}

\begin{document}

\begin{titlepage}

\begin{center}

\today\hfill
hep-th/0401139\\

\vskip 2 cm
{\large \bf Boundary RG Flows of N=2 Minimal Models}
\vskip 1 cm 
{Kentaro Hori}\\
\vskip 0.5cm
{\it 
University of Toronto,
Toronto, Ontario, Canada}\\

\end{center}

\vskip 0.5 cm
\begin{abstract}
We study boundary renormalization group flows of
${\mathcal N}=2$ minimal models using Landau-Ginzburg description of B-type.
A simple algebraic relation of matrices is relevant. 
We determine the pattern of the flows and identify
the operators that generate them.
As an application, we show that the charge lattice
of B-branes in the level $k$ minimal model
is $\Z_{k+2}$. We also reproduce the fact that
the charge lattice for the A-branes is
$\Z^{k+1}$, applying the B-brane analysis on the mirror LG orbifold.
\end{abstract}

\end{titlepage}

\newpage

\section{Introduction}\label{sec:intro}

Many systems in statistical mechanics and quantum field theory
have effective description of Landau-Ginzburg (LG) type.
In particular, in $(2,2)$ supersymmetric
field theories in $1+1$ dimensions, LG models
provide effective description of a large class of theories, both
conformal and massive, from which 
one can extract intuitive pictures as well as exact results, such as
the set and character of vacua, dimension of operators and chiral rings.
The simplest example is the single variable model with superpotential
\beq
W=X^{k+2}
\label{Wmin}
\eeq
labeled by a positive integer $k$.
It flows in the infra-red limit to the ${\mathcal N}=2$
minimal model at level $k$ ---
a $(2,2)$ superconformal field theory with central charge
$c={3k\over k+2}$ \cite{ZF,Qiu1,Ademollo,Yang,Qiu2},
as argued in \cite{Martinec,VWa,WiLG}.
LG description gives us a clear picture of renormalization group (RG)
flows between conformal field theories.
The system with superpotential (\ref{Wmin}) has 
left and right $U(1)$ R-symmetries which become a part of
the superconformal algebra. A vector $U(1)$ is lost 
if we add a lower order term
\beq
W=X^{k+2}+\varepsilon X^{\ell}.
\label{bulkpert}
\eeq
This can be regarded as a supersymmetric perturbation of
the minimal model by a relevant operator,
and one can immediately tell by looking at the superpotential
that it flows to the model of lower level $\ell-2$.

In this paper, we study the boundary RG flows of Landau-Ginzburg models
with an unbroken ${\mathcal N}=2$ supersymmetry.
Boundary RG flows
are being studied from two view points ---
statistical mechanics of two-dimensional critical systems
and string theory.
In the latter, boundary RG flows describe, from the worldsheet 
perspective,
the tachyon condensation on the worldvolume of
unstable D-brane systems \cite{HKM}.
The subject of unstable D-brane systems \cite{Senreview}
has proved to be extremely rich: 
it motivated the development of
string field theory \cite{SFT,BSFT},
led to the K-theory characterization of D-brane charge \cite{WK-theory}
and its refinement \cite{Douglas}, gave us a physical interpretation
of the matrix models \cite{Matrix},
and provided workable models of time dependent string theory
\cite{SenRoll}.
In most of them, an important role is played by
Chan-Paton factors which are
simple matrix factors that live on the worldsheet boundary.

\begin{figure}[htb]
\centerline{\includegraphics{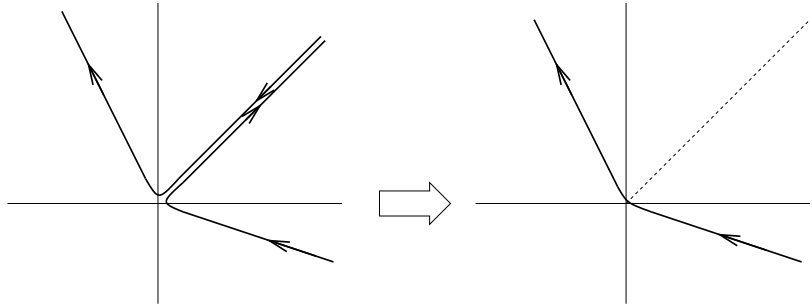}}
\caption{Flow of A-type boundary conditions in the minimal model}
\label{Fig:intro}
\end{figure}

A useful LG description of boundary RG flow already exists.
This is in the context of {\it A-branes}
 which are wrapped on Lagrangian submanifolds and support flat gauge fields.
In the minimal model and its deformations,
the branes are D1-branes at the
wedge-shaped lines that reside in the pre-image of
$W\in \R$ \cite{HIV}.
(The coset construction provides a similar and sometimes
useful picture where the branes are straight segment in a disc
stretched between special points on the boundary circle \cite{MMS}.)
This LG description provides a useful and geometrical picture of
RR-charge, Witten index, as well as the appearance of Verlinde algebra
\cite{HIV}.
In this description, the boundary RG flow is simply
the annihilation of the brane and antibrane (see Figure~\ref{Fig:intro}),
or recombination of the branes at the intersection points.
One can find the pattern of RG flows at a glance.
This has been remarked, for example, in \cite{Kennaway} (and also in the
disc picture in \cite{MMS,FreSch}).
However, it is not easy to identify the
operator that generates a given flow.

There are other class of branes, {\it B-branes},
which are wrapped on complex submanifolds and support
holomorphic bundles.
The purpose of this work is to find a useful picture of boundary RG flows
using B-branes in Landau-Ginzburg models, hopefully to the same extent as
the A-branes or even to complement what is missing in the A-type picture.
B-branes in LG models have been studied in
\cite{Hlin,HKKPTVVZ,HICM,KapLi,Orlov,Herbst,Lararoiu}.
In particular, we use the recent description by Kontsevich
\cite{KapLi,Orlov,Herbst,Lararoiu}
(an independent and alternative description is in
\cite{HICM})
that uses the factorization of the superpotential
on the Chan-Paton factor.
What is relevant here, it turns out, is the
continuous deformation of the matrices\footnote{The author learned this
in  \cite{Atiyah} where it is used 
in the proof of Bott periodicity.}
$$
M_t=\left(\begin{array}{cc}
A&0\\
0&1
\end{array}\right)
\left(\begin{array}{cc}
\cos t&-\sin t\\
\sin t&\cos t
\end{array}\right)
\left(\begin{array}{cc}
1&0\\
0&B
\end{array}\right)
\left(\begin{array}{cc}
\cos t&\sin t\\
-\sin t&\cos t
\end{array}\right)
$$
that yields the flow
\beq
\left(\begin{array}{cc}
A&0\\
0&B
\end{array}\right)
\,\mbox{\LARGE $\Rightarrow$}\,
\left(\begin{array}{cc}
AB&0\\
0&1
\end{array}\right).
\label{basic}
\eeq
This simple algebraic relation
provides the B-type counterpart of the A-type flow
as in Figure~\ref{Fig:intro}.
Also, by a basis change, $M_t\to M_t{\cos t\,\,-\sin t\choose
\sin t \,\,\,\,\cos t}$,  the matrix for small
$t$ can be written as
\beq
M_{\varepsilon}\cong \left(\begin{array}{cc}
A&0\\
0&B
\end{array}\right)
+\varepsilon \left(\begin{array}{cc}
0&-AB\\
1&0
\end{array}\right).
\label{bdrypert}
\eeq
This leads to the boundary analog of (\ref{bulkpert})
where the perturbing term breaks the $U(1)$ R-symmetry
and generates a flow of the boundary condition.
In this way, one can identify the operator that generates
a given flow, as well as identify the IR limit of a given perturbation.

As an application, we determine the charge lattice of the D-branes
in the minimal model.
For B-branes, it turns out that the lattice is torsion
$$
\Lambda_B\,\cong\,\Z_{k+2}.
$$
This is obtained through the relation (\ref{basic})
or more explicitly
$$
\left(\begin{array}{cccc}
x&&&\\
&\ddots&&\\
&&\ddots&\\
&&&x
\end{array}
\right)
\,\mbox{\LARGE $\Rightarrow$}\,
\left(\begin{array}{cccc}
x^{k+2}&&&\\
&1&&\\
&&\ddots&\\
&&&1
\end{array}
\right).
$$
We also reproduce the charge lattice of A-branes which has been known 
by the (A-type) LG picture as
$$
\Lambda_A\,\cong\,\Z^{k+1}.
$$
This is done by using the mirror symmetry between the minimal model and its
$\Z_{k+2}$-orbifold, where A and B are exchanged,
and applying (\ref{basic}) to the latter.

{\it Note:}
While this work is being written, we noticed a paper \cite{cappelli}
which studies boundary RG flows of non-supersymmetric minimal models
in LG description. Also, a paper \cite{Diaco} just appeared
which has some overlaps in the discussion of B-branes in LG orbifolds.

\section{B-branes in Landau-Ginzburg Models}\label{sec:BLG}

Let $X$ be a non-compact Calabi-Yau manifold with a (local)
K\"ahler potential $K(\phi,\overline{\phi})$
and a global holomorphic function $W(\phi)$.
We consider the $(2,2)$ supersymmetric LG model with the action
\beq
S=S_K+S_W=\int_{\Sigma}\dd^2x\left[\,
\int\dd^4\theta\, K(\Phi,\overline{\Phi})
\,+{\rm Re}\int\dd^2\theta\,W(\Phi)\,\right].
\eeq
For superspace and superfields
we use the convention of \cite{HKKPTVVZ}.
We are interested in B-branes of this system. Namely,
the boundary conditions and interactions
preserving the B-type ${\mathcal N}=2$ supersymmetry
$Q=Q_++Q_-$, $\oQ=\oQ_++\oQ_-$ \cite{OOY}.

Under the transformation
$\delta=\epsilon Q-\bepsilon \oQ$, 
the K\"ahler potential term (D-term) $S_K$ is invariant with the
ordinary supersymmetric boundary condition
for D-branes wrapped on a complex submanifold of $X$.
On the other hand,
the superpotential term (F-term) 
varies as \cite{Warner,Hlin}
\beq
\delta S_W
={\rm Re}\int_{\partial\Sigma,B}\dd t\,\dd\theta (-i\bepsilon)\,
W(\Phi)\,,
\eeq
where $\int_{\partial\Sigma,B}$ is the integration on the B-boundary
in which $\theta^+=\theta^-=:\theta$ (see \cite{Hlin} for conventions
on ``boundary superspace'').
This vanishes
 if the D-brane lies in a level set of the superpotential
$W$ \cite{GJS,HIV}.
There is actually an alternative way to preserve the B-type supersymmetry
\cite{Kachru}.
Suppose the superpotential can be written as the product
\beq
W(\Phi)=if(\Phi)g(\Phi).
\label{factorization}
\eeq
Then, we add a boundary term
\beq
S_{\rm bdry}=\int_{\partial\Sigma}\dd t\,\left[
\,{1\over 2}\int_B\dd\theta\dd\btheta\,
\overline{\Gamma}\Gamma\,+\,
{\rm Re}\int_B\dd\theta
\,\Gamma f(\Phi)\,
\right].
\eeq
where $\Gamma$ is a fermionic superfield on the B-boundary
which fails to be chiral,
\beq
\overline{D}\Gamma=g(\Phi).
\eeq
Under the B-type supersymmetry the boundary term varies as
\beqa
\delta S_{\rm bdry}
&=&{\rm Re}\int_{\partial\Sigma,B}\dd t\dd\theta \,
\left(\left[\epsilon{\mathcal Q}-\bepsilon\overline{\mathcal Q}\right]
\Gamma f(\Phi)\right)\Bigl|_{\btheta=0}
\nn\\
&=&{\rm Re}\int_{\partial\Sigma,B}\dd t\dd\theta \,
(-\bepsilon\overline{D}\Gamma f(\Phi))\Bigl|_{\btheta=0}
={\rm Re}\int_{\partial\Sigma,B}\dd t\dd\theta \,
i\bepsilon W(\Phi)\,,
\nn
\eeqa
which indeed cancels $\delta S_W$.
Thus we find a B-brane for each factorization of the superpotential
(\ref{factorization}).
More generally,
if the superpotential is expressed as
$W=i\sum_{\alpha=1}^ra_{\alpha}b_{\alpha}$, 
one can do the same by introducing $\Gamma_{\alpha}$
obeying $\overline{D}\Gamma_{\alpha}=b_{\alpha}(\Phi)$ with
the boundary superpotential
$\sum_{\alpha=1}^r\Gamma_{\alpha}a_{\alpha}(\Phi)$.
It is straightforward to generalize this to the case of gauged
linear or non-linear sigma models with superpotential.

The component expression of a superfield with constraint
$\overline{D}\Gamma
=g(\Phi)$ is $\Gamma=\eta+\theta G-i\theta\btheta\dot{\eta}
-\btheta g(\Phi)$ and the boundary term reads as
\beqa
S_{\rm bdry}&=&\int_{\partial\Sigma}\dd t\left\{
i\overline{\eta}\dot{\eta}-{1\over 2}|g(\phi)|^2
-{\rm Re}\Bigl(\overline{\eta}\partial_ig(\phi)\psi^i\Bigr)+{1\over 2}|G|^2
+{\rm Re}\Bigl(-\eta \partial_if(\phi)\psi^i +Gf(\phi)\Bigr)
\right\}
\nn\\
&=&\int_{\partial\Sigma}\dd t\left\{
i\overline{\eta}\dot{\eta}-{1\over 2}|g(\phi)|^2-{1\over 2}|f(\phi)|^2
-{\rm Re}\Bigl(\overline{\eta}\partial_ig(\phi)\psi^i
+\eta \partial_if(\phi)\psi^i\Bigr)
\right\}
\nn
\eeqa
where $\psi^i:=\psi^i_++\psi^i_-$.
The auxiliary field $G$ is eliminated by solving
$G=-\overline{f(\phi)}$.
The supersymmetry variation of the fermion $\eta$ is
$\delta\eta=\epsilon G-\bepsilon g(\phi)
=
-\epsilon \overline{f(\phi)}-\bepsilon g(\phi)$.
Let us formulate the system on the segment
$0\leq\s\leq\pi$ where we put
the fermions $\eta_{0}$ and $\eta_{\pi}$ at the two boundaries
with the boundary terms corresponding to
the factorizations
$$
W=if_0g_0=if_{\pi}g_{\pi}.
$$
The boundary at $\sigma=0$ is oriented toward the past $t\to-\infty$
while the $\sigma=\pi$
boundary is oriented toward the future $t\to+\infty$.
By N\"other procedure we find the
supercharge
$Q=Q_{\rm bulk}+Q_{\rm bdry}$,
$\oQ=\oQ_{\rm bulk}+\oQ_{\rm bdry}$
 where
\beqa
&&
\oQ_{\rm bulk}
=\int_{0}^{\pi}\dd\s \Bigl\{(\opsi_++\opsi_-)\cdot\partial_t\phi
+(\opsi_+-\opsi_-)\cdot\partial_{\s}\phi
+i(\psi_--\psi_+)^i\partial_iW(\phi)\Bigr\}
\nn\\
&&
\oQ_{\rm bdry}
=-i\Bigl[\,\eta f(\phi) \,+\, \overline{\eta}g(\phi)\,
\Bigr]^{\pi}_{0}.
\nn
\eeqa
Using the canonical (anti)commutation relation, we find
\beq
\oQ_{\rm bulk}^2=-iW\Bigl|_0^{\pi},
\quad
\{\oQ_{\rm bulk},\oQ_{\rm bdry}\}=0,
\quad
\oQ_{\rm bdry}^2=iW\Bigl|_0^{\pi},
\eeq
and indeed the total supercharge $\oQ$ is nilpotent.
The fermion at $\sigma=\pi$ is represented on a two dimensional vector
space $\C^2_{\pi}$
spanned by $|0\rangle_{\pi},\overline{\eta_{\pi}}|0\rangle_{\pi}$.
This is the Chan-Paton factor of the brane.
The Chan-Paton factor for the brane at $\s=0$ is likewise
$\C^2_0$ spanned by
$|0\rangle_{0},\overline{\eta_{0}}|0\rangle_{0}$.
Both spaces are $\Z_2$ graded by the fermion number ---
$|0\rangle$ is bosonic and $\overline{\eta}|0\rangle$ is fermionic.
Open string states take values in
$\Hom(\C_0^2,\C_{\pi}^2)$ on which the boundary part of the supercharge
is represented as
\beq
i\oQ_{\rm bdry}\left(\begin{array}{cc}
a&b\\
c&d
\end{array}\right)
=\left(\begin{array}{cc}
0&f_{\pi}(\phi)\\
g_{\pi}(\phi)&0
\end{array}\right)
\left(\begin{array}{cc}
a&b\\
c&d
\end{array}\right)
-\left(\begin{array}{cc}
a&-b\\
-c&d
\end{array}\right)
\left(\begin{array}{cc}
0&f_{0}(\phi)\\
g_{0}(\phi)&0
\end{array}\right).
\nn\\
\label{Qbdry}
\eeq
Since $X$ is assumed to be Calabi-Yau,
the supersymmetric ground states are determined by the zero mode
analysis. 
The zero mode Hilbert space is represented as
the space of 
$\Hom(\C^2_0,\C^2_{\pi})$-valued
antiholomorphic forms on $X$
on which the supercharge acts as the Dolbeault operator
plus (\ref{Qbdry}):
$$
i\oQ=\bartial+i\oQ_{\rm bdry}
$$
(In the zero mode sector,
the left and the right boundaries are mapped to the same point and thus
$\oQ_{\rm bulk}^2=\oQ_{\rm bdry}^2=0$.)
Note that this forms a $\Z_2$-graded complex where the grading comes from
the mod 2 reduction of form degree
and the fermion number in $\Hom(\C^2_0,\C^2_{\pi})$.
The space of supersymmetric ground states can be identified
as the $\oQ$-cohomology group.
If $X$ is a Stein space, this reduces to the cohomology of
$\oQ_{\rm bdry}$ acting on the space of
$\Hom(\C^2_0,\C^2_{\pi})$-valued holomorphic functions on $X$:
\beq
{\mathcal H}_{\rm SUSY}\cong
H(\Gamma_{\it hol}(X,\Hom(\C^2_0,\C^2_{\pi}));\oQ_{\rm bdry}).
\eeq

If the brane corresponds to more general
``factorization'', $W=i\sum_{\alpha=1}^{r}a_{\alpha}b_{\alpha}$,
the Chan-Paton factor is $\C^{2^r}$ spanned by the exterior
powers of $\overline{\eta}_{\alpha}$ multiplied to the state
$|0\rangle$ annihilated by $\eta_{\alpha}$. The operator $i\oQ_{\rm bdry}$
acts as $\sum_{\alpha=1}^r(\eta_{\alpha}a_{\alpha}+
\overline{\eta}_{\alpha}b_{\alpha})$, exchanging the bosonic and fermionic
subspaces of $\C^{2^r}$.
Let $f$ (resp. $g$) be the restriction of this operator
on the fermionic (resp. bosonic) subspace.
Then, $fg$ and $gf$ are both proportional to $-iW$.
One can further generalize the Chan-Paton factor to an arbitrary
$\Z_2$-graded vector space $V=V_+\oplus V_-$
on which there are operators $g:V_+\to V_-$ and
$f:V_-\to V_+$ such that
\beq
fg=-iW{\bf 1}_{V_+},\qquad
gf=-iW{\bf 1}_{V_-}.
\eeq
The boundary action is given by the super-Wilson-line
for the superconnection
$$
{\mathcal A}={1\over 2}\left(\begin{array}{cc}
ff^{\dag}+g^{\dag}g&\psi^i\partial_if+\opsi^{\bi}\partial_{\bi}g^{\dag}\\
\psi^i\partial_ig+\opsi^{\bi}\partial_{\bi}f^{\dag}&
f^{\dag}f+gg^{\dag}
\end{array}\right).
$$
This is the same as the standard
super-Wilson-line factor for the brane-antibrane system
\cite{Takayanagi,FinnPer} corresponding to
the tachyon field $T=f+g^{\dag}:V_-\to V_+$ (up to the shift by
${\rm Im}(W)$).
Let 
$(V_{-}\mathop{\rightleftharpoons}^{
\!\!\!\!\!\!f}_{\!\!\!\!\!\!g}V_{+})_0$ and
$(V_{-}\mathop{\rightleftharpoons}^{
\!\!\!\!\!\!f}_{\!\!\!\!\!\!g}V_{+})_{\pi}$
be two such branes and consider
the open string stretched between them.
The space of supersymmetric ground states
is isomorphic to the cohomology group
\beq
{\mathcal H}_{\rm SUSY}\cong
H(\Gamma_{\it hol}(X,\Hom(V_0,V_{\pi}));\oQ_{\rm bdry}),
\eeq
where $\oQ_{\rm bdry}$ is defined as in (\ref{Qbdry}).

This realization of the space of supersymmetric ground states
was obtained by Kontsevich whose work is interpreted in the above form
in \cite{KapLi}. An independent work on the same subject is done by
the author \cite{Hlin,HKKPTVVZ,HICM}. In \cite{HICM} a different form of
the cohomological realization is obtained
(see also \cite{HKKPTVVZ} for a preliminary version which gives
the derivation).
We also note that some mathematical predictions of
these results combined with Mirror Symmetry \cite{HV}
were confirmed in \cite{OhCho,Cho}.

\section{B-Branes in ${\mathcal N}=2$ Minimal Models}\label{sec:Bmin}

Let us consider
the Landau-Ginzburg model of a single variable $X$ with the K\"ahler
potential $K=|X|^2$ and
superpotential
$$
W=iX^{k+2}.
$$
Since the superpotential is homogeneous, this model has a vector R-symmetry
$U(1)_V:X(\theta^{\pm})\to\e^{2i\alpha\over k+2}X(\e^{-i\alpha}\theta^{\pm})$,
in addition to the axial R-symmetry
 $U(1)_A:X(\theta^{\pm})\to X(\e^{\mp i\beta}\theta^{\pm})$.
There is also a discrete $\Z_{k+2}$ symmetry generated by
\beq
\qquad \quad
\g:X\to \omega X,\qquad
\omega:=\e^{2\pi i\over k+2}.
\eeq
The system flows in the infra-red limit to a $(2,2)$ sueprconformal 
field theory of central charge $c={3k\over k+2}$, called the
${\mathcal N}=2$ minimal model of level $k$.
The two R-symmetries of the LG model define the
$U(1)$ currents of the $(2,2)$ superconformal algebra.

\subsection{The B-branes $\Scr{B}_L$}

Let us apply the result of the previous section 
to this Landau-Ginzburg model.
The superpotential can be factorized as $W=ix^{L+1}\cdot x^{k+1-L}$.
Thus, for each $L=-1,0,...,k,k+1$, we find a B-brane $\Scr{B}_L$
given by the boundary action
$$
\int_{\partial\Sigma}\dd t\left[{1\over 2}\int_B\dd\theta\dd\btheta\,
\overline{\Gamma}\Gamma
\,+\,
{\rm Re}\int_B\dd\theta\,\Gamma X^{L+1}\right]
$$
where $\Gamma$ obeys the constraint $\overline{D}\Gamma=
X^{k+1-L}$.
The boundary term is invariant
under the $\Z_{k+2}$ discrete symmetry if we let $\g$ acts on the superfield
$\Gamma$ by
\beq
\g:\Gamma\to \omega^{-L-1}\Gamma.
\label{gonG}
\eeq
Also the vector R-symmetry is preserved under the transformation
\beq
U(1)_V:\Gamma(\theta)\to\e^{i{k-2L\over k+2}\alpha}
\Gamma(\e^{-i\alpha}\theta).
\label{U1G}
\eeq
If the brane $\Scr{B}_L$ defines a conformal boundary condition in the
IR limit,
we expect that $U(1)_V$ becomes a part of the
${\mathcal N}=2$ superconformal algebra.

\subsection{Supersymmetric Ground States}

Let us find the supersymmetric ground states of the open strings.
We first consider the string with the both ends at
$\Scr{B}_L$.
Using (\ref{Qbdry}), we find
$$
i\oQ_{\rm bdry}
\left(\begin{array}{cc}
a(x)&b(x)\\
c(x)&d(x)
\end{array}\right)
=\left(\begin{array}{cc}
x^{L+1}c(x)+b(x)x^{k+1-L}&x^{L+1}(d(x)-a(x))\\
x^{k+1-L}(a(x)-d(x))&x^{k+1-L}b(x)+c(x)x^{L+1}
\end{array}\right)
$$
It vanishes if and only if $a(x)=d(x)$ and
$x^{k+1-L}b(x)+c(x)x^{L+1}=0$.
Then, it is $\oQ_{\rm bdry}$-exact when
$a(x)$ is divisible by $x^{L+1}$ or $x^{k+1-L}$,
$b(x)$ is divisible by $x^{L+1}$ and $c(x)$ is divisible by
$x^{k+1-L}$. 
If $L=-1$ or $L=k+1$, any $\oQ_{\rm bdry}$-closed state is
$\oQ_{\rm bdry}$-exact and thus
there is no supersymmetric ground state.
For $0\leq L\leq k$, there are non-zero
$\oQ_{\rm bdry}$-cohomology classes represented by
\beq
|j\rangle^{b}_{LL}
=\left(\begin{array}{cc}
x^j&0\\
0&x^j
\end{array}\right),
\quad
|j\rangle^{f}_{LL}=
\left(\begin{array}{cc}
0&x^{L-j}\\
-x^{k-L-j}&0
\end{array}\right);
\qquad
j=0,1,...,\min\{L,k-L\}.
\eeq
Similarly,
for the string stretched from $\Scr{B}_{L_1}$ to
$\Scr{B}_{L_2}$, the $\oQ_{\rm bdry}$-cohomology group is non-trivial
for $0\leq L_1,L_2\leq k$ and the basis are represented by
\beqa
&&
|j\rangle^{b}_{L_1L_2}
=
\left(\begin{array}{cc}
\!x^{j-{L_1-L_2\over 2}}\!&0\\
0&\!x^{j+{L_1-L_2\over 2}}\!
\end{array}\right),
\quad
|j\rangle^{f}_{L_1L_2}=
\left(\begin{array}{cc}
0&\!x^{{L_1+L_2\over 2}-j}\!\\
\!\!-x^{k-{L_1+L_2\over 2}-j}\!&0
\end{array}\right);\quad
\label{susyg12}\\
&&\qquad\qquad\qquad
\mbox{$j={|L_1-L_2|\over 2},{|L_1-L_2|\over 2}+1,...,
\min\Bigl\{{L_1+L_2\over 2},k-{L_1+L_2\over 2}\Bigr\}$}.
\label{jrange}
\eeqa
In all cases, there are equal number of bosonic and fermionic supersymmetric
ground states. This in particular means that the Witten index vanishes
\beq
\Tr\!\!\mathop{}_{\Scr{B}_{L_1}\Scr{B}_{L_2}}
(-1)^F=0.
\label{Wvanish}
\eeq

\subsection*{\it $\Z_{k+2}$ action}

The $\Z_{k+2}$ symmetry $X\to \omega X$,
acts on the boundary fermion for the brane $\Scr{B}_L$ as
$\eta\to \omega^{-L-1}\eta$ according to (\ref{gonG}).
The action on the Chan-Paton factor is determined
by the action on the ground state $|0\rangle_L$, which depends on
a choice
\beq
\rho_M(\g):|0\rangle_L\to \omega^{-{M+L+1\over 2}}|0\rangle_L
\label{defrhoM}
\eeq
parametrized by a mod $2(k+2)$ integer $M$ such that $L+M+1$ is even.
The action on the ground states for the open string stretched
from $(\Scr{B}_{L_1},\rho_{M_1})$ to $(\Scr{B}_{L_2},\rho_{M_2})$
is
\beq
\g:|j\rangle_{L_1L_2}^b\mapsto \omega^{j+{M_1-M_2\over 2}}
|j\rangle_{L_1L_2}^b,\quad
|j\rangle_{L_1L_2}^f\mapsto \omega^{-j-1+{M_1-M_2\over 2}}
|j\rangle_{L_1L_2}^f.
\label{gactjLL}
\eeq
The open string Witten index twisted by the symmetry $\g^m$ is given by
\beq
\Tr\!\!\mathop{}_{(\Scr{B}_{L_1},\rho_{M_1})
(\Scr{B}_{L_2},\rho_{M_2})}
(-1)^F\g^m
=\omega^{m{M_1-M_2\over 2}}
\sum_{j\in(\ref{jrange})}\Bigl(\omega^{mj}-\omega^{-m(j+1)}\Bigr).
\label{twW}
\eeq

\subsection*{\it R-charges}

Let us next analyze the R-charges of these supersymmetric ground states.
By (\ref{U1G}), the R-symmetry acts on the boundary fermion for
the brane $\Scr{B}_L$ as
$\eta\to\lambda^{k-2L}\eta$ where $\lambda=\e^{i\alpha\over k+2}$.
We let the Chan-Paton state $|0\rangle_L$ to transform as
$|0\rangle_L\to\lambda^{{k\over 2}-L}|0\rangle_L$ so that the two states
$|0\rangle_L$ and $\overline{\eta}|0\rangle_L$ have the opposite charges.
This yields the following R-transformation of the
$L_1$-$L_2$ open string states:
$$
U(1)_V:\left(\begin{array}{cc}
a(x)&b(x)\\
c(x)&d(x)
\end{array}\right)
\longrightarrow
\lambda^{\bullet}
\left(\begin{array}{cc}
\lambda^{-L_2+L_1}a(\lambda^2x)&\lambda^{-L_2+k-L_1}b(\lambda^2x)\\
\lambda^{L_2-k+L_1}c(\lambda^2x)&\lambda^{L_2-L_1}d(\lambda^2x)
\end{array}\right)
$$
where $\lambda^{\bullet}$ is the factor to be determined.
The ground states transform as
$|j\rangle_{L_1L_2}^b\mapsto \lambda^{2j+\bullet}|j\rangle_{L_1L_2}^b$
and $|j\rangle_{L_1L_2}^f\mapsto \lambda^{k-2j+\bullet}|j\rangle_{L_1L_2}^f$.
We now require that the charge spectrum to be symmetric
under the sign flip $q\to -q$.
This fixes $\lambda^{\bullet}$ to be $\lambda^{-{k\over 2}}$.
Then, we find that the R-charges of $|j\rangle_{L_1L_2}^b$ and
$|j\rangle_{L_1L_2}^f$ are
\beq
q_{|j\rangle^b_{L_1L_2}}=-q_{|j\rangle^f_{L_1L_2}}
={2j\over k+2}-{k\over 2(k+2)}.
\label{Rstate}
\eeq

\subsection{Boundary Chiral Primaries}

Just as for closed string, there is a one-to-one correspondence with
the open string supersymmetric ground states and
the boundary chiral ring elements:
\beq
\Scr{O}^b_{j,L_1L_2}\longleftrightarrow |j\rangle_{L_1L_2}^b,\qquad
\Scr{O}^f_{j,L_1L_2}\longleftrightarrow |j\rangle_{L_1L_2}^f,
\eeq
Their R-charges are obtained from (\ref{Rstate}) by the spectral flow
$q\to q+{c\over 6}$:
\beq
q_{\Scr{O}^b_{j,L_1L_2}}={2j\over k+2},\qquad
q_{\Scr{O}^f_{j,L_1L_2}}={k-2j\over k+2}
\label{Rbcr}
\eeq
Under the assumption that $\Scr{B}_L$ define conformal boundary conditions,
the boundary chiral ring elements define boundary chiral primary fields
of the boundary CFT. As usual \cite{LVW}, the R-charge $q$ determines
the conformal dimension of the operator $h={q\over 2}$.
Thus, the operators $\Scr{O}^{b,f}_{j,L_1L_2}$ have dimension
$h_{j,L_1L_2}^b={j\over k+2}$ and $h_{j,L_1L_2}^f={{k\over 2}-j\over k+2}$.

For the boundary preserving operators, $L_1=L_2=L$, they can be
expressed in terms of the elementary fields as
\beq
\Scr{O}^b_{j,LL}=x^j,\qquad
\Scr{O}^f_{j,LL}=\eta x^{L-j}-\overline{\eta} x^{k-L-j}.
\eeq
They indeed represent the non-trivial cohomology classes of
$\overline{\delta}x=0,\overline{\delta}\eta=-\bepsilon x^{k+1-L}$
and $\overline{\delta}\overline{\eta}=-\bepsilon x^{L+1}$,
and have the right R-charge (\ref{Rbcr}) under
$x\to \e^{2i\alpha\over k+2}x$,
$\eta\to \e^{{k-2L\over k+2}i\alpha}\eta$,
$\overline{\eta}\to \e^{-{k-2L\over k+2}i\alpha}\overline{\eta}$.
The fermionic operator $\Scr{O}_{j,LL}^f$ is the lowest component of
the superfield
$\Gamma X^{L-j}-\overline{\Gamma} X^{k-L-j}$ which is indeed chiral
since $\overline{D}\Gamma=X^{k+1-L}$ and
$\overline{D}\overline{\Gamma}=X^{L+1}$ (on shell).
Thus, the corresponding deformation is given by the boundary F-term
$$
{\mit\Delta}S_{\rm bdry}
={\rm Re}\int_{\partial\Sigma,B}\dd t\dd\theta\,
\Bigl(
\Gamma X^{L-j}-\overline{\Gamma} X^{k-L-j}\Bigr)
$$
The $t$ integrand is an operator of dimension
$$
{{k\over 2}-j\over k+2}+{1\over 2}={k+1-j\over k+2}<1,
$$
and therefore is relevant.

\subsection{The branes $\Scr{B}_{-1}$ and $\Scr{B}_{k+1}$}

The branes $\Scr{B}_{-1}$ and $\Scr{B}_{k+1}$ are special in the sense
that there is no supersymmetric ground state on the open strings stretched
between themselves as well as between them and any other brane.
Also, either $f$ or $g$ of the factorization of $W$ is 1 (identity)
and the term  $|f|^2+|g|^2$ of the boundary action is larger than
or equal to $1$
everywhere on the field space. This is analogous to the situation of
``constant tachyon'' where we expect that the brane decays to nothing.
From these facts, we claim that the branes
$\Scr{B}_{-1}$ and $\Scr{B}_{k+1}$ can be regarded as ``nothing''
or ``zero''.
Namely, if they make a summand of the brane,
like $\Scr{B}_{-1}\oplus \Scr{B}'$,
the boundary condition in the infra-red limit is identified as
$\Scr{B}'$ only.

\subsection{Comparison with RCFT results}

A detailed study of the boundary state
of ${\mathcal N}=2$ minimal model is done in a part of
\cite{BH2}, based on an earlier work \cite{MMS} which studies the
D-branes in the coset model ${SU(2)_k\times U(1)_2\over U(1)_{k+2}}$
using the standard RCFT technique \cite{Cardy}.
The coset model is obtained from the minimal model by a particular non-chiral
GSO projection, and what is done in \cite{BH2} is to carefully
identify the boundary state before that GSO projection.
It is found that there is a boundary condition $\Scr{B}_{L,M,S}$
labeled by $L\in\{0,1,...,k\}$, $M\in \Z_{2(k+2)}$ and
$S\in \Z_4$ with $L+M+S$ even, which preserves the
supersymmetry $Q_+-(-1)^SQ_-$ (thus $S$ odd ones are relevant for us).
It is invariant under the $\Z_{k+2}$ discrete symmetry
and the boundary state on the circle twisted
by $\gamma^m$ is\footnote{This slightly differs from the state
$|\Scr{B}_{[j,s]}\rangle$ in \cite{BH2}
by a phase, $|\Scr{B}_{[j,s]}\rangle_{a^{2m}}
=\e^{{\pi i nm\over k+2}-{\pi is\over 2}}|\Scr{B}_{2j,n,s}\rangle_{\g^m}$
and $|\Scr{B}_{[j,s]}\rangle_{(-1)^Fa^{2m}}
=\e^{{\pi i nm\over k+2}}|\Scr{B}_{2j,n,s}\rangle_{(-1)^F\g^m}$.}
\beqa
&&|\Scr{B}_{L,M,S}\rangle_{\g^{m}}
=(2k+4)^{{1\over 4}}\e^{-\pi i{Mm\over k+2}+\pi i {S\over 2}}
\sum_{l\in \{0,1,...,k\}\atop \nu\in \{0,1\}}
{S_{L\,l}\over\sqrt{S_{0\,l}}}(-1)^{S\nu}
\kket{l,m,2\nu+1}_B,\quad
\label{BRR}\\
&&|\Scr{B}_{L,M,S}\rangle_{(-1)^F\g^{m}}
=(2k+4)^{{1\over 4}}\e^{-\pi i{Mm\over k+2}}
\sum_{l\in \{0,1,...,k\}\atop \nu\in \{0,1\}}
{S_{L\,l}\over\sqrt{S_{0\,l}}}(-1)^{S\nu}
\kket{l,m,2\nu}_B,
\label{BNSNS}
\eeqa
where $\kket{j,m,s}_B$ is the B-type Ishibashi state.
From this one can read the full spectrum of open strings,
including the supersymmetric ground states with their $\Z_{k+2}$-charges.
In particular, the
$\Z_{k+2}$-twisted Witten index can be computed as the overlap
(see Eqn (5.43) of \cite{BH2}):
\beqa
\lefteqn{
{}_{\g^m}\langle \Scr{B}_{L_1,M_1,S_1}|\e^{\pi i J_0}q_t^H|
\Scr{B}_{L_2,M_2,S_2}\rangle_{\g^m}}
\nn\\
&&=\e^{\pi i\left({m(M_1-M_2)\over k+2}-{S_1-S_2\over 2}\right)}
\sum_{l\in \{0,1,...,k\}}
N_{L_1L_2}^l
\left(\e^{\pi i {ml\over k+2}}-\e^{-\pi i{m(l+2)\over k+2}}\right),
\eeqa
where $N_{l\,l'}^{l''}$ is the $SU(2)_k$ Fusion coefficients.
This agrees with (\ref{twW}) under the identification of
$(\Scr{B}_L,\rho_M)$ and $\Scr{B}_{L,M,1}$.
The NSNS part of the boundary state ($m=0$ in (\ref{BNSNS}))
shows that the boundary entropy \cite{gdef}
of the brane $\Scr{B}_L$ is
given by
\beq
g_L:=\langle 0|\Scr{B}_L\rangle_{\NSNS}
=(2k+4)^{1\over 4}{S_{L\,0}\over\sqrt{S_{0\,0}}}
=\sqrt{2}
{\sin(\pi {L+1\over k+2})\over\sqrt{\sin({\pi\over k+2})}}.
\label{bentr}
\eeq

We also need to stress that we do not have the so called
``short-orbit branes'' $\widehat{\Scr{B}}$
in our system. They are the oriented branes
in the GSO projected model (coset model) \cite{MMS},
where nothing is wrong with the coexistence of $\widehat{\Scr{B}}$
 and our branes.
In the model with the other GSO projection, opposite in the RR sector,
our branes become oriented and short-orbit branes are unoriented,
but again there is no problem with the coexistence.
However, in the model before the GSO projection,
there is an odd number of real fermions between our branes
$\Scr{B}_L$ and the short-orbit branes $\widehat{\Scr{B}}$,
which is problematic in quantization \cite{WK-theory}.
Thus our branes and short-orbit branes cannot coexist before GSO projection.
This fact is very important in the construction of rational B-branes
in Gepner model, directly from the B-branes in the minimal models
\cite{BHHW}. The LG description of the short-orbit branes
is given in the model with superpotential
$W=X^{k+2}-Y^2$. A related discussion is given
in the third paper of \cite{KapLi}.

\section{Brane Charges and RG Flows}\label{sec:Charge}

We have seen that the Witten index of the open string for any pair of B-branes
vanishes (\ref{Wvanish}). By factorization,
this implies that the overlap of the boundary state
and the RR ground states all vanish
${}_{\RR}\langle i|\Scr{B}_L\rangle_{\RR}=0$,
which is indeed the case: (\ref{BRR}) has no overlap with
the supersymmetric ground states
$|\ell,\ell+1,1\rangle\otimes |\ell,-\ell-l,-1\rangle$
on the untwisted circle.
However, this does not mean that the D-brane charge vanishes.
There could be a torsion charge that cannot be measured by the overlaps,
${}_{\RR}\langle i|\Scr{B}\rangle_{\RR}$.
In this section, we show that our B-branes $\Scr{B}_L$ with $L=0,1,...,k$
indeed carry  such torsion charges.
What we use extensively is the homotopy relation between matrices of
the type (\ref{basic}), 
$$
\left(\begin{array}{cc}
A&0\\
0&B
\end{array}\right)\,
\stackrel{0\leftarrow t}{\longleftarrow}
\,
R_t
\left(\begin{array}{cc}
A&0\\
0&B
\end{array}\right)
R_{t}^{-1}
\,
\stackrel{t\to {\pi\over 2}}{\longrightarrow}
\,
\left(\begin{array}{cc}
B&0\\
0&A
\end{array}\right)
$$
where
\beq
R_t:=\left(\begin{array}{cc}
\cos t&-\sin t\\
\sin t&\cos t
\end{array}\right).
\label{defRt}
\eeq
Furthermore, we determine the pattern of boundary RG flows
and identity the operators that generate them.

\subsection{The Charges of The B-branes}

Let us consider the one-parameter family of $2\times 2$ matrix pairs
\beqa
f_t(x)&=&
R_t
\left(\begin{array}{cc}
1&0\\
0&x
\end{array}\right)
R_t^{-1}
\left(\begin{array}{cc}
x&0\\
0&1
\end{array}\right)
\label{homotopy1}
\\
g_t(x)&=&
\left(\begin{array}{cc}
1&0\\
0&x
\end{array}\right)
R_t
\left(\begin{array}{cc}
x^{k+1}&0\\
0&x^k
\end{array}\right)R_t^{-1},
\label{homotopy2}
\eeqa
where $R_t$ is the matrix given by (\ref{defRt}).
One can readily see that the condition of B-type supersymmetry
is preserved at each $t$,
$f_tg_t=x^{k+2}{\bf 1}_2$,
$g_tf_t=x^{k+2}{\bf 1}_2$.
At $t=0$, the linear maps are
$$
f_0(x)=\left(\begin{array}{cc}
x&0\\
0&x
\end{array}\right),
\quad
g_0(x)=\left(\begin{array}{cc}
x^{k+1}&0\\
0&x^{k+1}
\end{array}\right),
$$
representing the sum of two copies of the $L=0$ brane,
$\Scr{B}_0\oplus\Scr{B}_0$.
At $t={\pi \over 2}$, the linear maps are
$$
f_{\pi \over 2}(x)=\left(\begin{array}{cc}
x^2&0\\
0&1
\end{array}\right),
\quad
g_{\pi \over 2}(x)=\left(\begin{array}{cc}
x^{k}&0\\
0&x^{k+2}
\end{array}\right),
$$
representing the sum of $L=1$ and $L=-1$ branes,
$\Scr{B}_1\oplus\Scr{B}_{-1}$.
Since the $L=-1$ brane is empty,
we find that the two brane configurations
are continuously connected with each other.
$$
\Scr{B}_0\oplus\Scr{B}_0\simeq
\Scr{B}_1.
$$
Similarly, by the following family of configurations
\beqa
f_t(x)&=&R_t
\left(\begin{array}{cc}
1&0\\
0&x^{L_2+1}
\end{array}\right)
R_t^{-1}
\left(\begin{array}{cc}
x^{L_1+1}&0\\
0&1
\end{array}\right),
\label{homot1}
\\
g_t(x)&=&
\left(\begin{array}{cc}
x^{k-L_1-L_2}&0\\
0&x^{k+1-L_2}
\end{array}\right)
R_t
\left(\begin{array}{cc}
x^{L_2+1}&0\\
0&1
\end{array}\right)R_t^{-1}
\label{homot2}
\eeqa
we find the homotopy relation of the branes
\beq
\Scr{B}_{L_1}\oplus\Scr{B}_{L_2}\,\simeq\,
\Scr{B}_{L_1+L_2+1}.
\label{sumrel}
\eeq
Here we have assumed that $L_1+L_2\leq k$.
If $L_1+L_2\geq k$,
we find $\Scr{B}_{L_1}\oplus\Scr{B}_{L_2}\,\simeq\,
\Scr{B}_{L_1+L_2-k-1}$ by a suitable modification of
the homotopy.
Using (\ref{sumrel}) repeatedly, we find
\beq
\Scr{B}_L\,\simeq\,
\underbrace{\Scr{B}_0\oplus\cdots\oplus\Scr{B}_0}_{L+1}.
\label{Bgen}
\eeq
As the spacial case $L=k+1$, we find
\beq
\underbrace{\Scr{B}_0\oplus\cdots\oplus\Scr{B}_0}_{k+2}
\,\simeq \,
0.
\label{Brel}
\eeq
The homotopy relations
(\ref{Bgen}) and (\ref{Brel})
imply that the RR-charge of the B-branes is
\beq
\Lambda_B\cong \Z_{k+2}
\eeq
generated by $\Scr{B}_0$.

\subsection{Mirror Picture: A-branes in the $\Z_{k+2}$ LG Orbifold}

The LG model with superpotential $W=X^{k+2}$
is mirror to the LG orbifold $\widetilde{W}=\widetilde{X}^{k+2}$ with respect
to the group $\Z_{k+2}$ \cite{Vafa} which acts
on the fields as $\widetilde{X}\to\omega \widetilde{X}$.
The B-branes in the model $W=X^{k+2}$
is mapped to the A-branes in the LG orbifold.
\begin{figure}[htb]
\centerline{\includegraphics{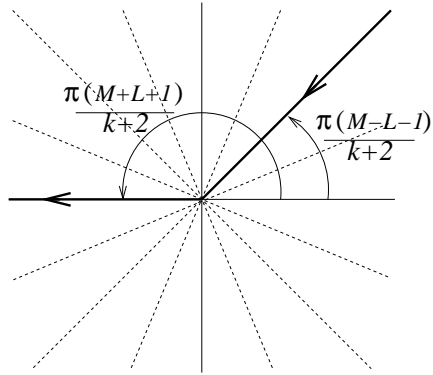}}
\caption{The A-brane $\Scr{A}_{L,M}$. This is the example
$L=2$, $M=5$ for $k=6$.}
\label{Abr}
\end{figure}
A-branes in the model before orbifold are
the D1-brane at the wedge-shaped lines with apex at $\widetilde{X}=0$,
which are mapped to the (positive) real line of
 the $\widetilde{W}$-plane.
For each $L=0,1,...,k$ and $M\in \Z$ (mod $2(k+2)$) such that
$L+M+1$ is even, there is an A-brane $\Scr{A}_{L,M}$
which the wedge coming in from
the direction $\arg(\widetilde{x})=\pi{M-L-1\over k+2}$ and going out to
the direction $\arg(\widetilde{x})=\pi{M+L+1\over k+2}$.
See Figure~\ref{Abr}.
The opening angle and the mean-direction of the wedge
are determined by $L$ and $M$ respectively as
$2\pi{L+1\over k+2}$ and ${\pi M\over k+2}$.

The orbifold group $\Z_{k+2}$ acts on the branes as
$\Scr{A}_{L,M}\to \Scr{A}_{L,M+2}$ which is free of fixed points nor
stabilizers.
Thus A-branes in the orbifold model are just the sums over images.
We denote by $\Scr{A}_L$ the brane obtained from $\Scr{A}_{L,M}$
which have opening angle $2\pi{L+1\over k+1}$.
Let us consider the sum of two such branes
$\Scr{A}_{L_1}\oplus\Scr{A}_{L_2}$
which are obtained from the sums over images of
$\Scr{A}_{L_1,M_1}$ and $\Scr{A}_{L_2,M_2}$.
One can find a pair $(M_1,M_2)$ such that the out-going direction of
$\Scr{A}_{L_1,M_1}$ agrees with the in-coming direction of
$\Scr{A}_{L_2,M_2}$. Under such arrangement, the charge of the sum
$\Scr{A}_{L_1,M_1}\oplus \Scr{A}_{L_2,M_2}$ is the same as
the charge of the brane $\Scr{A}_{L_1+L_2+1,M'}$ for some $M'$.
\begin{figure}[htb]
\centerline{\includegraphics{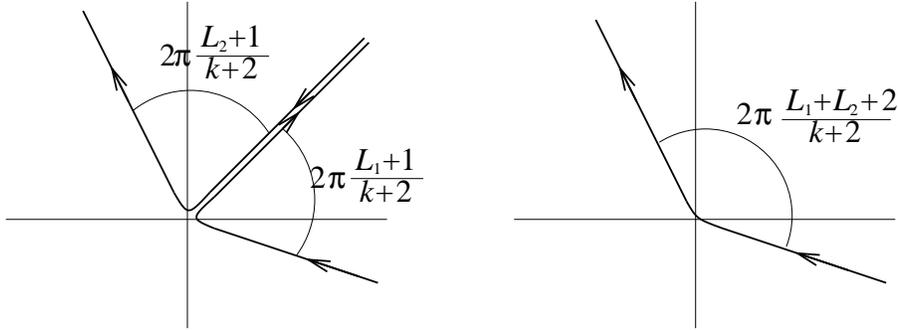}}
\caption{Cancellation of out-going and in-coming rays of two A-branes.}
\label{cancel}
\end{figure}
(We will discuss the A-brane charge in the Section~\ref{subsec:MA}
in more detail.)
This is understood as the cancellation of the two parallel
rays with the opposite orientations. See Figure~\ref{cancel}.
Thus, we find the homotopy relation
\beq
\Scr{A}_{L_1}\oplus\Scr{A}_{L_2}\,\simeq\,\Scr{A}_{L_1+L_2+1}.
\eeq
This is the mirror counterpart of the relation
(\ref{sumrel}) under the identification
\beq
\Scr{B}_{L}\,=\,\Scr{A}_L.
\eeq
In particular, the relation (\ref{Brel}) is mirror to the
process that the sum of $(k+2)$-copies of the brane
$\Scr{A}_0$ of opening angle ${2\pi\over k+2}$ annihilates to nothing,
by cancellation of the out-going ray of one brane and the in-coming ray
of the next brane.
See Figure~\ref{trivial} in Section~\ref{sec:LGorb}
for the corresponding annihilation
in the model before orbifold.

\subsection{Boundary RG Flows}\label{subsec:bRG}

The cancellation of the parallel rays of opposite orientations
we have seen is nothing but annihilation of brane and antibrane,
which can be regarded as a tachyon condensation on the worldvolume
\cite{Sen}.
In the worldsheet perspective, open string tachyon condensations
can be described as the boundary renormalization group flows
generated by boundary relevant operators \cite{HKM,Callan,PolTh,FSW}.
We would like to view the homotopy relation
$\Scr{B}_{L_1}\oplus\Scr{B}_{L_2}\to \Scr{B}_{L_1+L_2+1}$ as
such a boundary RG flow, directly for B-branes in the
model without orbifold,
and identify the relevant operator that generates it.

To this end, we make a basis change of the Chan-Paton factors 
that simplifies the expression 
of the intermediate configurations (\ref{homot1})-(\ref{homot2}). 
(Recall that we are assuming $L_1+L_2\leq k$. Other cases can be
treated with an obvious modification.)
We change the basis of $\C^2_+$ by
$R_t$, so that the matrix expression changes as
$f_t(x)\to R_t^{-1}f_t(x)$, $g_t(x)\to g_t(x)R_t$. More explicitly
we have
\beqa
&&f_t(x)=
\left(\begin{array}{cc}
1&0\\
0&\!x^{L_2+1}\!
\end{array}\right)
R_t^{-1}
\left(\begin{array}{cc}
\!x^{L_1+1}\!&0\\
0&1
\end{array}\right)
=\left(\begin{array}{cc}
x^{L_1+1}\cos t&\sin t\\
\!-x^{L_1+L_2+2}\sin t\!&x^{L_2+1}\cos t
\end{array}\right),
\nn\\
&&g_t(x)=
\left(\begin{array}{cc}
\!\!x^{k-L_1-L_2}\!\!&0\\
0&\!\!\!x^{k+1-L_2}\!\!
\end{array}\right)
R_t
\left(\begin{array}{cc}
\!x^{L_2+1}\!&0\\
0&1
\end{array}\right)
=
\left(\begin{array}{cc}
x^{k+1-L_1}\cos t&\!-x^{k-L_1-L_2}\sin t\!\\
x^{k+2}\sin t&x^{k+1-L_2}\cos t
\end{array}\right)
\nn
\eeqa
In this expression, we see that the configuration for
$t=\varepsilon \ll 1$ can be expanded as
\beqa
&&f_{\varepsilon}(x)=
f_0(x)+\varepsilon \left(\begin{array}{cc}
0&1\\
-x^{L_1+L_2+2}&0
\end{array}\right)+\cdots,
\label{pert1}
\\
&&g_{\varepsilon}(x)=
g_0(x)+\varepsilon\left(\begin{array}{cc}
0&-x^{k-L_1-L_2}\\
x^{k+2}&0
\end{array}\right)+\cdots.
\label{pert2}
\eeqa
To identify the operator that gives this perturbation,
we note that the matrix entries of $f(x)$ and $g(x)$ have the following 
invariant meaning:
$$
f(x)=\left(\begin{array}{cc}
|0\rangle_{\!\!{}_1}{}_{{}_1\!\!}\langle 0|\eta_1&
|0\rangle_{\!\!{}_1}{}_{{}_2\!\!}\langle 0|\eta_2\\
|0\rangle_{\!\!{}_2}{}_{{}_1\!\!}\langle 0|\eta_1&
|0\rangle_{\!\!{}_2}{}_{{}_2\!\!}\langle 0|\eta_2
\end{array}\right),\quad
g(x)=\left(\begin{array}{cc}
\bareta|0\rangle_{\!\!{}_1}{}_{{}_1\!\!}\langle 0|&
\bareta_1|0\rangle_{\!\!{}_1}{}_{{}_2\!\!}\langle 0|\\
\bareta_2|0\rangle_{\!\!{}_2}{}_{{}_1\!\!}\langle 0|&
\bareta_2|0\rangle_{\!\!{}_2}{}_{{}_2\!\!}\langle 0|
\end{array}\right).
$$
This is enough to find that the perturbation
corresponds to the following fermionic states
in the $\Scr{B}_{L_1}$-$\Scr{B}_{L_2}$ sector and
the $\Scr{B}_{L_2}$-$\Scr{B}_{L_1}$ sector
\beq
|{\rm pert}\rangle^f_{\!{}_{L_1L_2}}=
\left(\begin{array}{cc}
0&-x^{L_1+L_2+2}\\
x^{k+2}&0
\end{array}\right),
\quad
|{\rm pert}\rangle^f_{\!{}_{L_2L_1}}=
\left(\begin{array}{cc}
0&1\\
-x^{k-L_1-L_2}&0
\end{array}\right)
\eeq
Comparing with (\ref{susyg12}),
these states can be formally identified as the
``supersymmetric ground states'' $-|j_{12}\rangle_{L_1L_2}^f$
and $|j_{21}\rangle^f_{L_2L_1}$ respectively, where
$j_{12}=-{L_1+L_2\over 2}-2$ and
$j_{21}={L_1+L_2\over 2}$.
Note that $j_{21}$ is in the range 
(\ref{jrange}) but $j_{12}$ is not.
Thus, $|{\rm pert}\rangle^f_{L_2L_1}=|j_{21}\rangle_{L_2L_1}^f$
is a true ground state
but $|{\rm pert}\rangle^f_{L_1L_2}=-|j_{12}\rangle_{L_1L_2}^f$
is not.
In fact,
$|{\rm pert}\rangle^f_{L_1L_2}$ is fine in the sense that it is
annihilated by
the supercharge $\oQ_{\rm bdry}$ but it is a
$\oQ_{\rm bdry}$-exact state.
We can interpret what we have seen as follows:
{\it The perturbation (\ref{pert1})-(\ref{pert2}) is generated by the
boundary F-term corresponding to the fermionic state}
\beq
|j_{21}\rangle_{L_2L_1}^f;\qquad
j_{21}={L_1+L_2\over 2}.
\label{perte}
\eeq
{\it In order to define a deformation
which obeys the supersymmetry condition,
$f(x)g(x)=-iW$ and $g(x)f(x)=-iW$, it needs to be accompanied by
a $\oQ$-exact part, which is
$-|j_{12}\rangle_{L_1L_2}^f$, $j_{12}=-{L_1+L_2\over 2}-2$.}

\begin{figure}[htb]
\centerline{\includegraphics{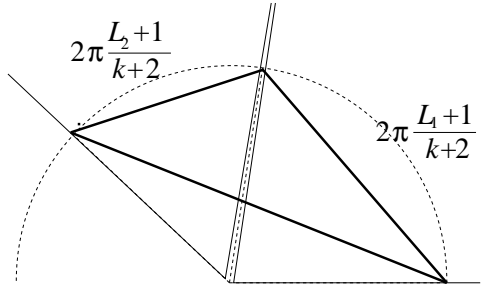}}
\caption{The disc picture of the mirror A-branes. The boundary entropy
is given by the length of the segment. It decreases under the RG flow
$\Scr{B}_{L_1}\oplus\Scr{B}_{L_2}\to\Scr{B}_{L_1+L_2+1}$ by
the triangle inequality.}
\label{dpic}
\end{figure}

It is conjectured that
the boundary entropy $g=\langle 0|\Scr{B}\rangle_{{}_{\rm NSNS}}$
must decrease under the boundary RG flows ---
``g-theorem'' \cite{gdef,gtheorem}.
Let us check this in the present case.
The boundary entropy of the brane $\Scr{B}_{L}$ is given by
(\ref{bentr}) or
$$
g_L=c_k\cdot \sin\left(\pi{L+1\over k+2}\right),
$$
where $c_k$ is an $L$-independent constant.
A nice picture to understand it is the one regarding the
${\mathcal N}=2$ minimal model or its orbifold as the dilatonic
sigma model on the disc where the A-branes are given by
the straight segments connecting $k+2$ special points on the
boundary \cite{MMS}.\footnote{However,
one can also
compute it in the LG picture using the proportionality
to the ``RR-charge'' $\langle 0|\Scr{B}\rangle_{{}_{\rm RR}}$
which is identified as the
weighted integral $\int_{L}\e^{-iW}\dd \widetilde{X}$ \cite{HIV}.
The disc picture
may be regarded
as focusing on the origin
$\widetilde{X}=0$ of the LG model in the IR limit.
The sharp apex of the wedge may be modified as the
straight segment of the disc (private communication with J.~Maldacena, 2002).}
The boundary entropy of a brane is proportional to the length
of the segment.
In this picture it is clear that the boundary entropy decreases under
the flow $\Scr{B}_{L_1}\oplus\Scr{B}_{L_2}\to
\Scr{B}_{L_1+L_2+1}$. It is simply the triangle inequality
(see Figure~\ref{dpic}).

\noindent
{\bf Remarks.}\\
(i)~Although $g_{L_1+L_2+1}$ is less than
the sum $g_{L_1}+g_{L_2}$, it is larger than the individual
entropy $g_{L_1}, g_{L_2}$. This is related to the fact that
we need to add the $\oQ$-trivial piece $|{\rm pert}\rangle_{L_1L_2}^f$
whose corresponding operator formally has dimension
${k+1-j_{12}\over k+2}=1+{L_1+L_2+2\over 2(k+2)}$ larger than $1$.\\
(ii)~It would be a very interesting problem to define the boundary entropy
throughout the RG flow, not just the UV and IR limits,
and see if it continuously decreases.
See \cite{Kutasov} for a proposal in the context of 
supersymmetric field theories in (bulk) four-dimensions.

\subsubsection{More general perturbations}

The perturbation operator corresponding to (\ref{perte})
has dimension 
$$
\Delta_{\rm pert}={k+1-j_{21}\over k+2}
={k+1-{L_1+L_2\over 2}\over k+2}.
$$
It is the most relevant operator since the maximum value of $j$
for $|j\rangle_{L_1L_2}^f$ or
$|j\rangle^f_{L_2L_1}$ is
$j_{\rm max}={\rm min}\{{L_1+L_2\over 2},k-{L_1+L_2\over 2}\}$
(which is ${L_1+L_2\over 2}$
in the present case where $L_1+L_2\leq k$ is assumed).
We would now like to study the perturbation generated by other
relevant operators, $j={|L_1-L_2|\over 2},...,j_{\rm max}-1$.

\begin{figure}[htb]
\centerline{\includegraphics{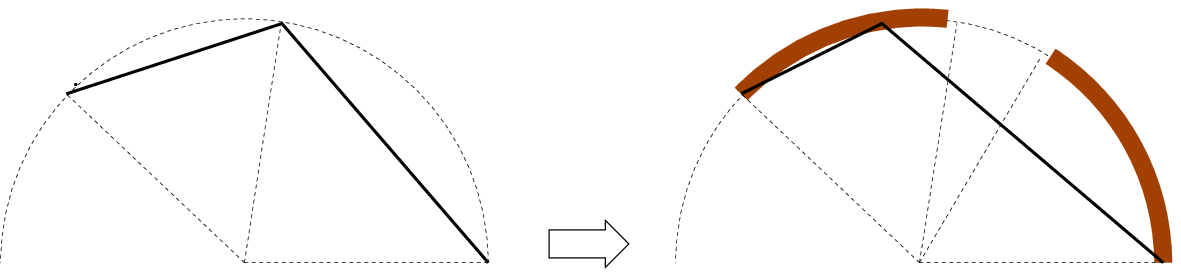}}
\caption{The RG flow $\Scr{B}_{L_1}\oplus\Scr{B}_{L_2}
\to \Scr{B}_{{L_1+L_2\over 2}+j+1}\oplus\Scr{B}_{{L_1+L_2\over 2}-j-1}$
generated by the operator corresponding to
$|j\rangle_{L_1L_2}^f$ or $|j\rangle_{L_2L_1}^f$.
The mid-point after the flow can only be in the shaded regions.
The two components correspond to whether the flow is generated by
$|j\rangle_{L_1L_2}^f$ or $|j\rangle_{L_2L_1}^f$.
It is evident that the sum of lengths decreases under the flow.}
\label{flow1}
\end{figure}

We consider the following generalization of
the family of configurations
\beqa
&&f_t(x)=
\left(\begin{array}{cc}
1&0\\
0&\!x^a\!
\end{array}\right)
R_t^{-1}
\left(\begin{array}{cc}
\!\!x^{L_1+1}\!\!&0\\
0&\!\!x^{L_2+1-a}\!\!
\end{array}\right)
=\left(\begin{array}{cc}
x^{L_1+1}\cos t&x^{L_2+1-a}\sin t\\
\!-x^{L_1+1+a}\sin t\!&x^{L_2+1}\cos t
\end{array}\right),
\label{ho1}\\
&&g_t(x)=
\left(\begin{array}{cc}
\!\!x^{k+1-L_1-a}\!\!&0\\
0&\!\!x^{k+1-L_2}\!\!
\end{array}\right)
R_t
\left(\begin{array}{cc}
\!x^a\!&0\\
0&1
\end{array}\right)
=
\left(\begin{array}{cc}
x^{k+1-L_1}\cos t&\!\!-x^{k+1-L_1-a}\sin t\!\\
x^{k+1-L_2+a}\sin t&x^{k+1-L_2}\cos t
\end{array}\right)
\nn\\
\label{ho2}
\eeqa
for some integer $a$ such that matrix entries of
the right hand sides are all monomials of $x$:
\beq
{\rm max}\{-L_1-1,\,L_2-k-1\}
\leq a\leq 
{\rm min}\{L_2+1,\,k+1-L_1\}.
\label{ine}
\eeq
As before we can identify this as the perturbation generated by
the operator corresponding to the fermionic states
$-|j_{12}\rangle^f_{L_1L_2}$ and $|j_{21}\rangle^f_{L_2L_1}$ where
$$
j_{12}={L_2-L_1\over 2}-a-1,\qquad
j_{21}={L_1-L_2\over 2}+a-1.
$$
Note that $j_{12}+j_{21}=-2$ and hence it is impossible for both
to satisfy the condition (\ref{jrange}) ---
at most only one of them can satisfy it. If we assume $L_1\geq L_2$ and
$L_1+L_2\leq k$ in addition to (\ref{ine}),
the condition is
\beqa
&&|j_{12}\rangle^f_{L_1L_2}\in {\mathcal H}_{\rm SUSY}
\,\,\Longleftrightarrow\,\,
a\leq L_2-L_1-1,
\label{ineqc1}\\
&&|j_{21}\rangle^f_{L_2L_1}\in {\mathcal H}_{\rm SUSY}
\,\,\Longleftrightarrow\,\,
a\geq 1.
\label{ineqc2}
\eeqa
Only if either one of these conditions are met,
can one consider $f_{\varepsilon}(x),g_{\varepsilon}(x)$ as
a relevant perturbation of $\Scr{B}_{L_1}\oplus \Scr{B}_{L_2}$.
The final configuration is the one with $t={\pi\over 2}$.
After the change of basis of $\C_+^2$
by $R_{\pi\over 2}={0\,-1\choose 1\,\,0}$,
it is expressed as
\beq
\widetilde{f}(x)=\left(\begin{array}{cc}
x^{L_1+1+a}&0\\
0&x^{L_2+1-a}
\end{array}\right),\quad
\widetilde{g}(x)=\left(\begin{array}{cc}
x^{k+1-L_1-a}&0\\
0&x^{k+1-L_2+a}
\end{array}\right),
\eeq
which is the configuration of the sum
$\Scr{B}_{L_1+a}\oplus\Scr{B}_{L_2-a}$.
Note that $\{L_1+a,L_2-a\}=\{{L_1+L_2\over 2}+j+1,{L_1+L_2\over 2}-j-1\}$
for $j=j_{12}$ or $j=j_{21}$.
Thus we find that the relevant operator corresponding to
$|j\rangle^f_{L_1L_2}\in {\mathcal H}_{\rm SUSY}$ or
$|j\rangle^f_{L_2L_1}\in {\mathcal H}_{\rm SUSY}$ generates an RG flow
\beq
\Scr{B}_{L_1}\oplus\Scr{B}_{L_2}
\,\,\longrightarrow\,\,
\Scr{B}_{{L_1+L_2\over 2}+j+1}\oplus\Scr{B}_{{L_1+L_2\over 2}-j-1}.
\eeq
Note that the difference of the two opening angles
$|L_1-L_2|$ increases after the flow
$|({L_1+L_2\over 2}+j+1)-({L_1+L_2\over 2}-j-1)|
=2j+2>|L_1-L_2|$.
In the disc picture of the mirror minimal model, we indeed see that
the total boundary entropy decreases (see Figure~\ref{flow1}):
\beq
g_{L_1}+g_{L_2}> g_{{L_1+L_2\over 2}+j+1}+g_{{L_1+L_2\over 2}-j-1}.
\eeq

Finally, we would like to comment on operators in
the $\Scr{B}_{L_i}$-$\Scr{B}_{L_i}$ sectors.
As we have seen, there are indeed fermionic
chiral primary operators that may correspond to relevant deformations.
However, one cannot find a deformation that obeys the
supersymmetry condition $f(x)g(x)=-i W$ and
$g(x)f(x)=-iW$. Thus, it seems that these operators cannot induce
a finite supersymmetric deformation of the system.
This guarantees the conservation of the torsion D-brane
charge $\Lambda_B\cong \Z_{k+2}$ we have claimed.

\section{B-Branes in LG Orbifold $W=X^{k+2}/\Z_{k+2}$}\label{sec:LGorb}

In this section, we study the charges and the boundary RG flows for
B-branes in the LG orbifold of $W=X^{k+2}$ with respect
to the group $\Z_{k+2}$ generated by $\g:X\to \omega X$.
This is mirror to A-branes in the LG model without orbifold
which are studied in \cite{HIV}. We first describe that known cases and 
then see how the result is reproduced and extended using
B-branes of the orbifold.

\subsection{Mirror picture: A-branes in $W=\widetilde{X}^{k+2}$}
\label{subsec:MA}

\begin{figure}[htb]
\centerline{\includegraphics{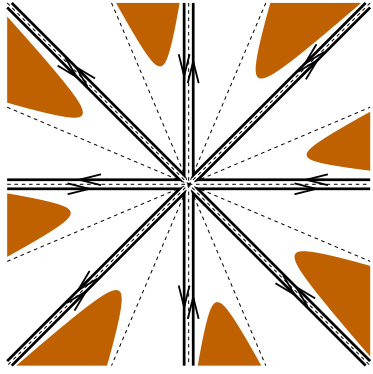}}
\caption{The branes $\Scr{A}_{0,M}$, $M=0,2,...,2(k+1)$ (for the case
$k=6$).
The shaded regions are $B^+$.
With a small positive rotation, each represent a class
$[\Scr{A}_{0,M}^+]\in H_1(\C,B^+)$.
The sum annihilates in $H_1(\C,B^+)$ because of the cancellation
of the out-going ray of one brane and the in-coming ray of the
next brane.}
\label{trivial}
\end{figure}

We recall that we have an A-brane $\Scr{A}_{L,M}$ in the LG model
$W=\widetilde{X}^{k+2}$ for each $(L,M)$ with $L+M=1$ is even. It 
is the wedge coming from $\arg(\widetilde{x})=\pi{M-L-1\over k+2}$ 
and going to
$\arg(\widetilde{x})=\pi{M+L+1\over k+2}$
(see Figure~\ref{Abr}).
The Witten index of the open string stretched between two of such branes
is
\beq
I(\Scr{A}_1,\Scr{A}_2)=\#(\Scr{A}_1^-\cap
\Scr{A}_2^+)
\eeq
where $\Scr{A}^{\pm}$ is the rotation of $\Scr{A}$ by
a small positive/negative angle (see \cite{HKKPTVVZ}).
The space of supersymmetric ground states of the string is
\beq
{\mathcal H}^b_{\rm SUSY}\oplus {\mathcal H}^f_{\rm SUSY}=\left\{
\begin{array}{ll}
\C\oplus 0&{\rm if}\,\,I=1\\
0\oplus \C&{\rm if}\,\,I=-1\\
0\oplus 0&{\rm if}\,\,I=0.
\end{array}\right.
\eeq
If $L_1\geq L_2$, $k\geq L_1+L_2$, the index is given by
\beq
I(\Scr{A}_{L_1,M_1},\Scr{A}_{L_2,M_2})
=\left\{
\begin{array}{ll}
\,\,1\,&{\rm if}\,\,\,{M_2-L_2-1\over 2}+1\leq {M_1+L_1+1\over 2}\leq
{M_2+L_2+1\over 2}\\[0.2cm]
\,-1\,&{\rm if}\,\,\,{M_2-L_2-1\over 2}+1\leq {M_1-L_1-1\over 2}\leq
{M_2+L_2+1\over 2}\\[0.2cm]
\,\,0&\mbox{otherwise}.
\end{array}\right.
\label{Aind}
\eeq
The charge lattice of the A-branes is $H_1(\C,B^+)$ where $B^+$ is the
region in which ${\rm Im}(W)$ is large positive.
It is
\beq
\Lambda_A\cong \Z^{k+1}
\eeq
generated by $[\Scr{A}^+_{0,M}]$ $M=0,2,...,2(k+1)$
which obey the linear relation
$$
[\Scr{A}^+_{0,0}]+[\Scr{A}^+_{0,2}]+\cdots+[\Scr{A}^+_{0,2k+2}]=0\qquad
{\rm in}\,\,\,H_1(\C,B^+).
$$
Figure~\ref{trivial} describes this relation in the example of $k=6$.

\subsubsection{RG-flow as brane recombination}\label{subsec:ARG}

Let us study the supersymmetric boundary RG flows.
We first consider a single brane $\Scr{A}_{L,M}$.
This represents a non-trivial charge $[\Scr{A}_{L,M}^+]\in H_1(\C,B^+)$
and must be stable.
Indeed, since $I(\Scr{A}_{L,M},\Scr{A}_{L,M})=1$,
 there is only one supersymmetric ground state and it is bosonic.
Therefore there is no fermionic chiral primary field and hence no
supersymmetric deformation operator.
We next consider the sum of two branes
$\Scr{A}_{L_1,M_1}\oplus\Scr{A}_{L_2,M_2}$.
Whether there is another brane configuration to which it can decay
depends on the intersection numbers of the two branes.
There are four cases to consider.

\noindent
(i) \underline{No intersection}\\
If the two do not intersect, there is no supersymmetric ground state, and
hence no supersymmetric deformation of the brane configuration,
both in the 1-2 and 2-1 string sectors.
Since there is neither in the 1-1 and 2-2 sectors (as we have seen for
the single brane case), there is no supersymmetric deformation of the
brane.
Thus the brane is stable.

\begin{figure}[htb]
\centerline{\includegraphics{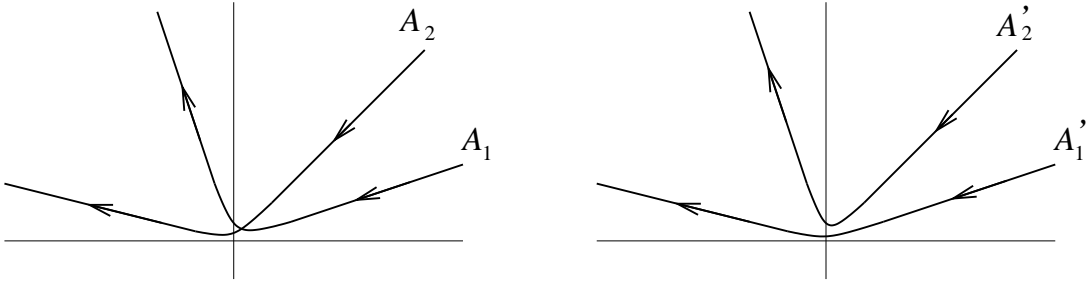}}
\caption{Two branes intersecting ``transversely'' (Left).
It is expected to recombine to two other branes without intersection
(Right).}
\label{int1}
\end{figure}

\noindent
(ii) \underline{``Transverse'' intersection}\\
This is the case where the two intersects transversely as in
Figure~\ref{int1}(Left).
In this case, one of the intersection numbers is $+1$ and other
is $-1$. In the Figure it is 
$I(1,2)=1$ and $I(2,1)=-1$.
In this case, there is one fermionic supersymmetric ground state
in the 2-1 string sector.
Depending on the dimension of the corresponding deformation operator,
the brane can decay into another brane configuration.
The wedge-picture suggests that the brane recombination occurs, and it ends
up with a configuration of two other branes.
The end point is as in Figure~
(Right) which is the sum
$\Scr{A}_{L_1',M_1'}\oplus\Scr{A}_{L_2',M_2'}$,
where $L_1'={L_1+L_2-M_1+M_2\over 2},M_1'={-L_1+L_2+M_1+M_2\over 2},$
$L_2'={L_1+L_2+M_1-M_2\over 2},M_2'={L_1-L_2+M_1+M_2\over 2}$.

\begin{figure}[htb]
\centerline{\includegraphics{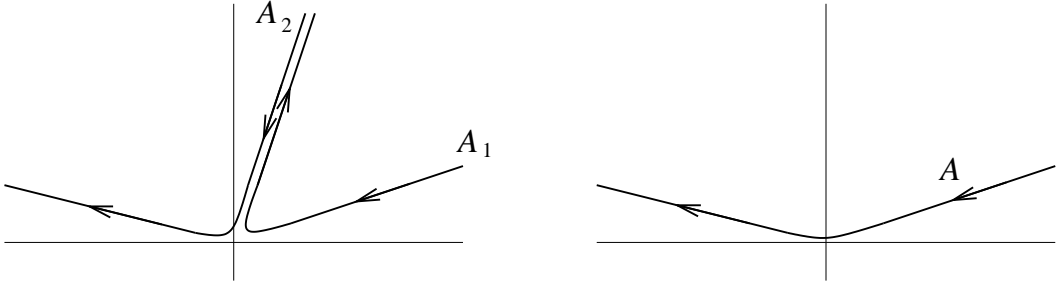}}
\caption{Two branes intersecting ``non-transversely'' (Left).
They are expected to combine into one brane (Right) by canceling the
parallel rays of opposite orientation.}
\label{int2}
\end{figure}

\noindent
(iii) \underline{``Non-transverse'' intersection, $-1$}\\
If the incoming ray of one wedge is the same as the out-going ray of the
other, one of the intersections is $0$ but the other is
$\pm 1$ depending on the orientations of the two branes.
Here we consider the $-1$ case, as in Figure~\ref{int2}(Left)
where $I(1,2)=0$ and $I(2,1)=-1$.
In this case, there is one fermionic supersymmetric ground state
from the 2-1 string sector.
If the dimension of the corresponding deformation operator is less than 1,
it can decay to a new brane configuration.
The wedge-picture suggests it is the brane obtained by
deleting the overlapping and oppositely oriented rays of the two branes.
For the case as in Figure where
$\pi{M_1+L_1+1\over k+2}=\pi{M_2-L_2-1\over k+2}$ (mod $2\pi$)
and $L_1+L_2\leq k$,
it is $\Scr{A}_{L,M}$ with
$L=L_1+L_2+1$ and
$M={-L_1+L_2+M_1+M_2\over 2}$.

\begin{figure}[htb]
\centerline{\includegraphics{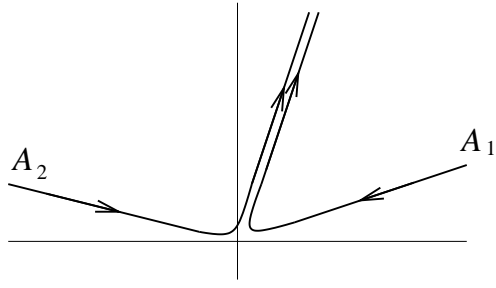}}
\caption{Two branes intersecting ``non-transversely''.
They cannot combine into one brane because the orientation of the parallel
rays are the same.}
\label{int3}
\end{figure}

\noindent
(iv) \underline{``Non-transverse'' intersection, $+1$}\\
Next, we consider the $+1$ case, as in Figure~\ref{int3}
where $I(1,2)=0$ and $I(2,1)=1$, which is obtained by flipping the
orientation of one of the two branes.
In this case, there is one bosonic ground state but
no fermionic ground state.
Thus, there is no supersymmetric deformation and the brane is stable.

\subsection{B-branes in the orbifold}

We now describe the same system in the mirror LG orbifold.

Let us denote by $\Scr{B}_{L,M}$ the B-brane in the orbifold
obtained from the B-brane $(\Scr{B}_{L},\rho_M)$ of the original model
where $\rho_M$ is the Chan-Paton representation of the orbifold group
$\Z_{k+2}$ given by (\ref{defrhoM}),
$$
\g:|0\rangle_L\mapsto \omega^{-{M+L+1\over 2}}|0\rangle_L,
\quad
\overline{\eta}|0\rangle_L\mapsto
\omega^{-{M-L-1\over 2}}\overline{\eta}|0\rangle_L.
$$
Note that the ``brane configuration''
$(\C\overline{\eta}|0\rangle_L
\mathop{\rightleftharpoons}^{\!\!\!\!\!\!f}_{\!\!\!\!\!\!g}\C|0\rangle_L)$
is invariant under the orbifold group
\beqa
&f(x)=x^{L+1}\to \omega^{-{M+L+1\over 2}}f(\omega x)
\omega{M-L-1\over 2}=f(x),
\nn\\
&g(x)=x^{k+1-L}\to
\omega^{-{M-L-1\over 2}}g(\omega x)\omega^{M+L+1\over 2}=g(x).
\nn
\eeqa

\subsubsection{Open string ground states}

Let us first analyze the supersymmetric ground states of the open string
stretched between two such branes, $\Scr{B}_{L_1,M_2}$ and
$\Scr{B}_{L_2,M_2}$.
First thing to note is that the $\Z_{k+2}$-equivariant
$\oQ$-cohomology is the same as the $\Z_{k+2}$-invariant states of the
ordinary $\oQ$-cohomology.
Let us explain this.
Let $(C,Q)$ be a complex and let $\Gamma$ be a symmetry group.
Namely $\Gamma$ acts on $C$ and each $\gamma\in \Gamma$
preserves the degree and commutes with $Q$.
Then, the $\Gamma$-invariant elements of $C$ form a complex $C^{\Gamma}$.
We assume $\Gamma$ is a compact group.
Then we have a

\noindent
{\bf Theorem.} {\it $\Gamma$-invariant
part of the cohomology group is the same as
the cohomology of $\Gamma$-invariant part of the complex}
\beq
H(C,Q)^{\Gamma}=H(C^{\Gamma},Q).
\eeq
\noindent
{\bf Proof:} The proof assumes that $\Gamma$ is finite but it is clear this
applies also to a group with an invariant measure with volume 1.
We have a map
\beq
Z^{\Gamma}\longrightarrow (Z/QC)^{\Gamma}.
\label{abovem}
\eeq
We want to show (i) it is surjective, and (ii) the Kernel is
$Q(C^{\Gamma})$. To show (i) let $f\in C$ represent a 
$\Gamma$-invariant cohomology
class. Namely, $Qf=0$ and, for any $\gamma\in \Gamma$,
$\gamma f=f+Qf'_{\gamma}$ for some
$f'_{\gamma}\in C$.
Then,
$$
f_1:={1\over |\Gamma|}\sum_{\gamma\in \Gamma}\gamma f=f+Qf'
$$
for some $f'\in C$. Namely, $f\equiv f_1\in Z^{\Gamma}$.
This shows (i).
To show (ii), let $f\in Z^{\Gamma}$ be mapped to $0$ by (\ref{abovem}).
Namely, $\gamma f=f (\forall \gamma\in \Gamma)$,
$Qf=0$, and $f=Qf'$ for some $f'\in C$.
Then,
$$
f={1\over |\Gamma|}\sum_{\gamma\in \Gamma}\gamma f
={1\over |\Gamma|}\sum_{\gamma\in \Gamma}\gamma 
Qf'=Q\left({1\over |\Gamma|}\sum_{\gamma\in \Gamma}\gamma f'\right).
$$
This means $f\in Q(C^{\Gamma})$. Thus (ii) is shown. {\bf End.}

The orbifold group action on the
supersymmetric ground states of the original theory
have been found in (\ref{gactjLL}).
There is at most one invariant state. It is $|j_*\rangle_{L_1L_2}^b$
or $|j_*'\rangle_{L_1L_2}^f$ if $j_*$ or $j_*'$ defined by
$$
j_*\equiv {M_2-M_1\over 2},\qquad
j_*'\equiv {M_1-M_2\over 2}-1\qquad\mbox{mod $(k+2)$},
$$
is in the range (\ref{jrange}), and there is none if both
of them are outside that range.
Thus, we find
\beq
{\mathcal H}^b_{\rm SUSY}\oplus {\mathcal H}^f_{\rm SUSY}=\left\{
\begin{array}{ll}
\C\oplus 0&{\rm if}\,\,
{|L_1-L_2|\over 2}\leq \Bigl[{M_2-M_1\over 2}\Bigr]_{k+2}\leq
\min\Bigl\{{L_1+L_2\over 2},k-{L_1+L_2\over 2}\Bigr\}\\[0.2cm]
0\oplus \C&{\rm if}\,\,
{|L_1-L_2|\over 2}\leq \Bigl[{M_1-M_2\over 2}-1\Bigr]_{k+2}\leq
\min\Bigl\{{L_1+L_2\over 2},k-{L_1+L_2\over 2}\Bigr\}
\\[0.2cm]
0\oplus 0&\mbox{otherwise},
\end{array}\right.
\label{susyg}
\eeq
where $[c]_{k+2}$ is the lowest non-negative number in
$c+(k+2)\Z$. 
This agrees with the ground state spectrum for the pair
$\Scr{A}_{L_1,M_1}$-$\Scr{A}_{L_2,M_2}$. For example, compare with
(\ref{Aind}) for the case $L_1\geq L_2$ and $k\geq L_1+L_2$.
Thus we find the following identification
\beq
\Scr{A}_{L,M}=\Scr{B}_{L,M}.
\eeq
What we learn from this is the dimension of the operator $\Scr{O}$
for the supersymmetric deformations.
The identification given in Section~\ref{sec:Bmin}
 of the R-charge of the ground states and corresponding
operators goes through without modification also in the orbifold theory.
In particular, the dimension of the operator that corresponds to the
fermionic supersymmetric ground state (present in the middle case of
(\ref{susyg})) is
\beq
\Delta={k+1-j_*'\over k+2}\equiv
1-{M_1-M_2\over 2(k+2)}\,\,\,\mbox{mod $(k+2)$}.
\label{rdim}
\eeq
It is indeed a relevant operator, $\Delta <1$.

\subsubsection{RG flows}

Let us consider the sum of two branes $\Scr{B}_{L_1,M_1}\oplus
\Scr{B}_{L_2,M_2}$. 
The Chan-Paton factor is given by
$\C^2_+\oplus\C^2_-$ where $\C^2_+$ is spanned by
$\{|0\rangle_{L_1},|0\rangle_{L_2}\}$ and
$\C^2_-$ is spanned by
$\{\overline{\eta}_1||0\rangle_{L_1},\overline{\eta}_2|0\rangle_{L_2}\}$
on which the orbifold group acts as
$$
\g_{+}
=\left(\begin{array}{cc}
\omega^{-{M_1+L_1+1\over 2}}&0\\
0&\omega^{-{M_2+L_2+1\over 2}}
\end{array}\right),\quad
\g_{-}
=\left(\begin{array}{cc}
\omega^{-{M_1-L_1-1\over 2}}&0\\
0&\omega^{-{M_2-L_2-1\over 2}}
\end{array}\right),
$$
respectively. The maps $f:\C^2_-\to\C^2_+$
and $g:\C^2_+\to\C^2_-$ 
are represented by the matrices
$$
f(x)=\left(\begin{array}{cc}
x^{L_1+1}&0\\
0&x^{L_2+1}
\end{array}\right),\quad
g(x)=\left(\begin{array}{cc}
x^{k+1-L_1}&0\\
0&x^{k+1-L_2}
\end{array}\right),
$$
which are indeed invariant under the orbifold group action,
$f(x)\to\g_{+}f(\omega x)\g_{-}^{-1}$,
$g(x)\to\g_{-}g(\omega x)\g_{+}^{-1}$.
We would like to identify the supersymmetric
deformation given by the relevant operator 
corresponding to $|j_*'\rangle_{L_1L_2}^f$
when it is present.
Let us try the family of configurations given by (\ref{ho1}) and (\ref{ho2}):
\beqa
f_t(x)&=&
\left(\begin{array}{cc}
x^{L_1+1}\cos t&x^{L_2+1-a}\sin t\\
-x^{L_1+1+a}\sin t&x^{L_2+1}\cos t
\end{array}\right),
\label{orbho1}\\
g_t(x)&=&
\left(\begin{array}{cc}
x^{k+1-L_1}\cos t&-x^{k+1-L_1-a}\sin t\\
x^{k+1-L_2+a}\sin t&x^{k+1-L_2}\cos t
\end{array}\right)
\label{orbho2}
\eeqa
where $a$ is an integer such that
\beq
{\rm max}\{-L_1-1,\,L_2-k-1\}
\leq a\leq 
{\rm min}\{L_2+1,\,k+1-L_1\}.
\label{ine2}
\eeq
In the present case, we need to make sure that the deformation is
invariant under the orbifold group $\Z_{k+2}$ which acts as
$f_t(x)\to\g_{+}f_t(\omega x)\g_{-}^{-1}$,
$g_t(x)\to\g_{-}g_t(\omega x)\g_{+}^{-1}$.
This requires the condition
\beq
\omega^{a}=\omega^{M_2+L_2-M_1-L_1\over 2}.
\label{a-b}
\eeq
If $j_*'$ is in the allowed range (\ref{jrange}),
one can solve (\ref{a-b}) for $a$ within (\ref{ine2}).
Then it is clear that
$f_{\varepsilon}(x), g_{\varepsilon}(x)$ can be regarded as
the perturbation by the operator corresponding to the state
$|j_*'\rangle_{L_1L_2}$ (cf. Section~\ref{subsec:bRG}).
The perturbed theory flows in the IR limit to the configuration 
with $t={\pi\over 2}$.
After the basis change of $\C^2_+$ by
$R_{\pi\over 2}={0\,-1\choose \! 1\,\,\,\,0}$, it is given by
\beqa
&&f'(x)=
\left(\begin{array}{cc}
x^{L_1+a+1}&0\\
0&x^{L_2-a+1}
\end{array}\right),\quad
g'(x)=
\left(\begin{array}{cc}
x^{k-L_1-a+1}&0\\
0&x^{k-L_2+a+1}
\end{array}\right)
\nn\\
&&
\g_{+}'=
\left(\begin{array}{cc}
\omega^{-{M_2+L_2+1\over 2}}&0\\
0&\omega^{-{M_1+L_1+1\over 2}}
\end{array}\right),\quad
\g_{-}'=\g_-.
\nn
\eeqa
This is the configuration for
$\Scr{B}_{L_1',M_1'}\oplus\Scr{B}_{L_2',M_2'}$ where
\beqa
L_1'=L_1+a\equiv \mbox{${L_1+L_2-M_1+M_2\over 2}$ (mod $k+2$)},
\quad
M_1'=\mbox{${-L_1+L_2+M_1+M_2\over 2}$},
\label{L1M1'}\\
L_2'=L_2-a\equiv \mbox{${L_1+L_2+M_1-M_2\over 2}$ (mod $k+2$)},
\quad
M_2'=\mbox{${L_1-L_2+M_1+M_2\over 2}$}.
\label{L2M2'}
\eeqa
Thus, we found that the deformation by the operator corresponding to
$|j_*'\rangle_{L_1L_2}^f$ generates the RG flow
\beq
\Scr{B}_{L_1,M_1}\oplus\Scr{B}_{L_2,M_2}\,\longrightarrow\,
\Scr{B}_{L_1',M_1'}\oplus\Scr{B}_{L_2',M_2'}
\eeq
where the two sets of labels are related by (\ref{L1M1'}) and
(\ref{L2M2'}). The decrease of the boundary entropy can be shown in
the same way as in Section~\ref{subsec:bRG}.
This process is the mirror of the brane-recombination of the A-branes
as described in Case (ii) of Section~\ref{subsec:ARG}, see Figure~\ref{int1}.
This is enough to see that
the brane charge is generated by
$\Scr{B}_{0,M}$ with $M=0,2,...,2(k+1)$ which are related by
$$
[\Scr{B}_{0,0}]+[\Scr{B}_{0,2}]+\cdots +[\Scr{B}_{0,2k+2}]=0.
$$
The mirror of Case (iii) where two branes
$\Scr{A}_{L_1,M_1}\oplus\Scr{A}_{L_2,M_2}$ combine into one
$\Scr{A}_{L,M}$
is described by the flow with $a=L_2+1$.
The $\Z_{k+2}$-invariance condition (\ref{a-b}) in this case is nothing but
$\pi {M_1+L_1+1\over k+2}\equiv \pi {M_2-L_2-1\over k+2}$
(mod $2\pi$), which is the condition that the
out-going ray of $\Scr{A}_{L_1,M_1}$ agrees with the
in-coming ray of $\Scr{A}_{L_2,M_2}$.

\section*{Acknowledgement}

The author thanks J.~Walcher for many useful discussion including
motivating ones at the initial stage.
He also thanks A.~Kapustin, C.~Lazaroiu and W.~Lerche for discussions.
He thanks the KITP, Santa Barbara for hospitality where some part of
this work is done.
This work is supported in part by Connaught Foundation, NSERC
and the Alfred P. Sloan Foundation.


\begin{thebibliography}{99}

\small
\parskip=0pt plus 2pt

\bibitem{ZF}
A.~B.~Zamolodchikov and V.~A.~Fateev,
``Disorder fields in two-dimensional
conformal quantum field theory and N=2 extended supersymmetry'',
Sov.\ Phys.\ JETP {\bf 63} (1986) 913
[Zh.\ Eksp.\ Teor.\ Fiz.\  {\bf 90} (1986) 1553].

\bibitem{Qiu1}
Z.~Qiu,
``Nonlocal current algebra and
N=2 superconformal field theory in two-dimensions'',
Phys.\ Lett.\ B {\bf 188} (1987) 207.





\bibitem{Ademollo}
M.~Ademollo {\it et al.},
``Supersymmetric strings and color confinement'',
Phys.\ Lett.\ B {\bf 62} (1976) 105;
``Dual string with U(1) color symmetry'',
Nucl.\ Phys.\ B {\bf 111} (1976) 77.


\bibitem{Yang}
F.~Ravanini and S.~K.~Yang,
``Modular invariance in N=2 superconformal field theories'',
Phys.\ Lett.\ B {\bf 195} (1987) 202.

\bibitem{Qiu2}
Z.~Qiu,
``Modular invariant partition functions
for N=2 superconformal field theories'',
Phys.\ Lett.\ B {\bf 198} (1987) 497.



\bibitem{Martinec}
E.~J.~Martinec,
``Algebraic Geometry And Effective Lagrangians,''
Phys.\ Lett.\ B {\bf 217} (1989) 431;
``Criticality, Catastrophes And Compactifications,''
In Brink, L. (ed.) et al.: {\it
Physics and mathematics of strings, V.G. Knizhnik memorial vol.} 389-433. 

\bibitem{VWa}
C.~Vafa and N.~P.~Warner,
``Catastrophes And The Classification Of Conformal Theories,''
Phys.\ Lett.\ B {\bf 218} (1989) 51.

\bibitem{WiLG}
E.~Witten,
``On the Landau-Ginzburg description of N=2 minimal models,''
Int.\ J.\ Mod.\ Phys.\ A {\bf 9} (1994) 4783
[hep-th/9304026].



\bibitem{HKM}
J.~A.~Harvey, D.~Kutasov and E.~J.~Martinec,
``On the relevance of tachyons,''
hep-th/0003101.


\bibitem{Senreview}
A.~Sen,
``Non-BPS states and branes in string theory,''
hep-th/9904207.


\bibitem{SFT}
A.~Sen and B.~Zwiebach,
``Tachyon condensation in string field theory,''
JHEP {\bf 0003} (2000) 002
[hep-th/9912249].

\bibitem{BSFT}
A.~A.~Gerasimov and S.~L.~Shatashvili,
``On exact tachyon potential in open string field theory,''
JHEP {\bf 0010} (2000) 034
[hep-th/0009103];

D.~Kutasov, M.~Marino and G.~W.~Moore,
``Some exact results on tachyon condensation in string field theory,''
JHEP {\bf 0010} (2000) 045
[hep-th/0009148].

\bibitem{WK-theory}
E.~Witten,
``D-branes and K-theory,''
JHEP {\bf 9812} (1998) 019
[hep-th/9810188].

\bibitem{Douglas}
M.~R.~Douglas,
``D-branes, categories and N = 1 supersymmetry,''
J.\ Math.\ Phys.\  {\bf 42} (2001) 2818
[hep-th/0011017].

\bibitem{Matrix}
J.~McGreevy and H.~Verlinde,
``Strings from tachyons: The c = 1 matrix reloaded,''
JHEP {\bf 0312} (2003) 054
[hep-th/0304224];
 
I.~R.~Klebanov, J.~Maldacena and N.~Seiberg,
``D-brane decay in two-dimensional string theory,''
JHEP {\bf 0307} (2003) 045
[hep-th/0305159].


\bibitem{SenRoll}
A.~Sen,
``Rolling tachyon,''
JHEP {\bf 0204} (2002) 048
[hep-th/0203211];

M.~Gutperle and A.~Strominger,
``Spacelike branes,''
JHEP {\bf 0204} (2002) 018
[hep-th/0202210];

N.~Lambert, H.~Liu and J.~Maldacena,
``Closed strings from decaying D-branes,''
hep-th/0303139.


\bibitem{HIV}
K.~Hori, A.~Iqbal and C.~Vafa,
``D-branes and mirror symmetry,''
hep-th/0005247.


\bibitem{MMS}
J.~M.~Maldacena, G.~W.~Moore and N.~Seiberg,
``Geometrical interpretation of D-branes in gauged WZW models,''
JHEP {\bf 0107} (2001) 046
[hep-th/0105038].

\bibitem{Kennaway}
K.~D.~Kennaway,
``A geometrical construction of Recknagel-Schomerus boundary states in
linear sigma models,''
Nucl.\ Phys.\ B {\bf 647} (2002) 471
[hep-th/0203266].

\bibitem{FreSch}
S.~Fredenhagen and V.~Schomerus,
``D-branes in coset models,''
JHEP {\bf 0202} (2002) 005
[hep-th/0111189];
``On boundary RG-flows in coset conformal field theories,''
Phys.\ Rev.\ D {\bf 67} (2003) 085001
[hep-th/0205011];


S.~Fredenhagen,
``Organizing boundary RG flows,''
Nucl.\ Phys.\ B {\bf 660} (2003) 436
[hep-th/0301229].



\bibitem{Hlin}
K.~Hori,
``Linear models of supersymmetric D-branes,''
hep-th/0012179.



\bibitem{HKKPTVVZ}
K.~Hori, S.~Katz, A.~Klemm, R.~Pandharipande, R.~Thomas,
C.~Vafa, R.~Vakil and E.~Zaslow,
{\it Mirror Symmetry}, (AMS-CMI, 2003).


\bibitem{HICM}
K.~Hori,
``Mirror symmetry and quantum geometry,''
hep-th/0207068.


\bibitem{KapLi}
A.~Kapustin and Y.~Li,
``D-branes in Landau-Ginzburg models and algebraic geometry,''
JHEP {\bf 0312} (2003) 005
[hep-th/0210296];

``Topological correlators in Landau-Ginzburg models with boundaries,''
hep-th/0305136;

``D-branes in topological minimal models: The Landau-Ginzburg approach,''
hep-th/ 0306001.


\bibitem{Orlov}
D.~Orlov,
``Triangulated categories of singularities and D-branes in  Landau-Ginzburg
models,''
math.ag/0302304.

\bibitem{Herbst}
I.~Brunner, M.~Herbst, W.~Lerche and B.~Scheuner,
``Landau-Ginzburg realization of open string TFT,''
hep-th/0305133.

\bibitem{Lararoiu}
C.~I.~Lazaroiu,
``On the boundary coupling of topological Landau-Ginzburg models,''
hep-th/0312286.




\bibitem{Atiyah}
M.~F.~Atiyah, {\it K-theory}
Benjamin, New York, 1967.

\bibitem{cappelli}
A.~Cappelli, G.~D'Appollonio and M.~Zabzine,
``Landau-Ginzburg description of boundary critical phenomena in two
dimensions,''
hep-th/0312296.

\bibitem{Diaco}
S.~A.~Ashok, E.~Dell'Aquila and D.~E.~Diaconescu,
``Fractional Branes in Landau-Ginzburg Orbifolds,''
hep-th/0401135.


\bibitem{OOY}
H.~Ooguri, Y.~Oz and Z.~Yin,
``D-branes on Calabi-Yau spaces and their mirrors,''
Nucl.\ Phys.\ B {\bf 477} (1996) 407
[hep-th/9606112].


\bibitem{GJS}
S.~Govindarajan, T.~Jayaraman and T.~Sarkar,
``Worldsheet approaches to D-branes on supersymmetric cycles,''
Nucl.\ Phys.\ B {\bf 580} (2000) 519
[hep-th/9907131].



\bibitem{Warner}
N.~P.~Warner,
``Supersymmetry in boundary integrable models,''
Nucl.\ Phys.\ B {\bf 450} (1995) 663
[hep-th/9506064].



\bibitem{Kachru}
S.~Hellerman, S.~Kachru, A.~E.~Lawrence and J.~McGreevy,
``Linear sigma models for open strings,''
JHEP {\bf 0207} (2002) 002
[hep-th/0109069].


\bibitem{Takayanagi}
T.~Takayanagi, S.~Terashima and T.~Uesugi,
``Brane-antibrane action from boundary string field theory,''
JHEP {\bf 0103} (2001) 019
[hep-th/0012210].

\bibitem{FinnPer}
P.~Kraus and F.~Larsen,
``Boundary string field theory of the DD-bar system,''
Phys.\ Rev.\ D {\bf 63} (2001) 106004
[hep-th/0012198].


\bibitem{HV}
K.~Hori and C.~Vafa,
``Mirror Symmetry,''
hep-th/0002222.


\bibitem{OhCho}
C.-H.~Cho and Y.-G.~Oh,
``Floer cohomology and disc instantons of Lagrangian torus fibers
in Fano toric manifolds,''
math.SG/0308225.


\bibitem{Cho}
C.-H.~Cho,
``Holomorphic disc, spin structures and Floer cohomology
of the Clifford torus,''
math.SG/0308224.


\bibitem{LVW}
W.~Lerche, C.~Vafa and N.~P.~Warner,
``Chiral Rings In N=2 Superconformal Theories,''
Nucl.\ Phys.\ B {\bf 324} (1989) 427.




\bibitem{BH2}
I.~Brunner and K.~Hori,
``Orientifolds and mirror symmetry,''
hep-th/0303135.



\bibitem{Cardy}
J.~L.~Cardy,
``Boundary Conditions, Fusion Rules And The Verlinde Formula,''
Nucl.\ Phys.\ B {\bf 324} (1989) 581.


\bibitem{gdef}
I.~Affleck and A.~W.~W.~Ludwig,
``Universal Noninteger 'Ground State Degeneracy' In Critical Quantum
Systems,''
Phys.\ Rev.\ Lett.\  {\bf 67} (1991) 161.


\bibitem{BHHW}
I.~Brunner, K.~Hori, K.~Hosomichi and J.~Walcher,
``Orientifolds of Gepner Models'', hep-th/0401137.

\bibitem{Vafa}
C.~Vafa,
``String Vacua And Orbifoldized L-G Models,''
Mod.\ Phys.\ Lett.\ A {\bf 4} (1989) 1169.


\bibitem{Sen}
A.~Sen,
``Tachyon condensation on the brane antibrane system,''
JHEP {\bf 9808} (1998) 012
[hep-th/9805170].

\bibitem{Callan}
C.~G.~Callan, I.~R.~Klebanov, A.~W.~W.~Ludwig and J.~M.~Maldacena,
``Exact solution of a boundary conformal field theory,''
Nucl.\ Phys.\ B {\bf 422} (1994) 417
[hep-th/9402113].

\bibitem{PolTh}
J.~Polchinski and L.~Thorlacius,
``Free Fermion Representation Of A Boundary Conformal Field Theory,''
Phys.\ Rev.\ D {\bf 50} (1994) 622
[hep-th/9404008].

\bibitem{FSW}
P.~Fendley, H.~Saleur and N.~P.~Warner,
``Exact solution of a massless scalar field with a relevant boundary
interaction,''
Nucl.\ Phys.\ B {\bf 430} (1994) 577
[hep-th/9406125].



\bibitem{gtheorem}
I.~Affleck and A.~W.~W.~Ludwig,
``Exact conformal-field-theory results on the multichannel Kondo effect:
Single-fermion Green's function, self-energy, and resisitivity,''
Phys.\ Rev.\ B {\bf 48} (1993) 7297.




\bibitem{Kutasov}
D.~Kutasov,
``New results on the 'a-theorem' in four dimensional supersymmetric field
theory,''
hep-th/0312098.



\end{thebibliography}
\end{document}